\documentclass[aps,pre,noshowpacs,twocolumn]{revtex4}
\usepackage{bm}
\usepackage{graphicx}
\usepackage{mathtools}
\usepackage{color,soul}
\renewcommand{\nu}{{\upsilon}}

\usepackage{changes}
\renewcommand{\P}{{ P}}

\newcommand{\U}{{U}}

\newcommand{\diss}{{\text diss}}
\newcommand{\EQ}{\begin{equation}}
\newcommand{\EE}{\end{equation}}
\newcommand{\EQA}{\begin{eqnarray}}
\newcommand{\EEA}{\end{eqnarray}}

\usepackage{hyperref}
\usepackage[11pt]{moresize}

\usepackage{chngcntr}

\begin{document}
\title{ Optimal evolutionary decision-making to store immune memory}

\author{Oskar H Schnaack${}^{1,2}$}
\author{Armita Nourmohammad${}^{1,2,3}$\footnote{correspondence should be addressed to Armita Nourmohammad: armita@uw.edu}}

\affiliation{1. Max Planck Institute for Dynamics and Self-organization, Am Fa\ss berg 17, 37077 G\"ottingen, Germany\\
2. Department of Physics, University of Washington, 3910 15th Ave Northeast, Seattle, WA 98195\\
3. Fred Hutchinson Cancer Research Center, 1100 Fairview ave N, Seattle, WA 98109\\}

\begin{abstract}
The adaptive immune system provides a diverse set of molecules that can mount specific responses against a multitude of pathogens. Memory is a key feature of adaptive immunity, which allows organisms to respond more readily upon re-infections. However, differentiation of memory cells is still one of the least understood cell fate decisions. Here, we introduce a mathematical framework to characterize optimal strategies to store memory to maximize the utility of immune response over an organism's lifetime. We show that memory production should be actively regulated to balance between affinity and cross-reactivity of immune receptors for an effective protection against evolving pathogens. Moreover, we predict that specificity of memory should depend on the organism's lifespan, and shorter-lived organisms with fewer pathogenic encounters should store more cross-reactive memory. Our framework provides a baseline to gauge the efficacy of immune memory in light of an organism's coevolutionary history with pathogens.
\end{abstract}

\maketitle
\noindent {\small {\bf Keywords}: mmune memory, adaptive immune system, non-equilibrium decision-making, evolutionary optimization}\\\\

\section{Introduction}
Adaptive immunity in vertebrates develops during the lifetime of an organism to battle a multitude of evolving pathogens.  The central actors in our adaptive immune system are diverse B- and T-cells, whose unique surface receptors are generated through genomic rearrangement, mutation, and selection~\cite{Janeway:H7fnIHBf}. The diversity of receptors allows the immune system to mount specific responses against diverse pathogens. B-cell receptors (BCRs) in particular can specialize through a process of affinity maturation, which is a form of {\em somatic Darwinian evolution} within an individual to enhance the affinity of BCRs to pathogens. Several rounds of somatic mutation and selection during affinity maturation can increase binding affinities of BCRs up to 10,000 fold~\cite{Victora:2012gx,MeyerHermann:2012ja}.

\begin{figure*}[t!]
\includegraphics[width=\textwidth]{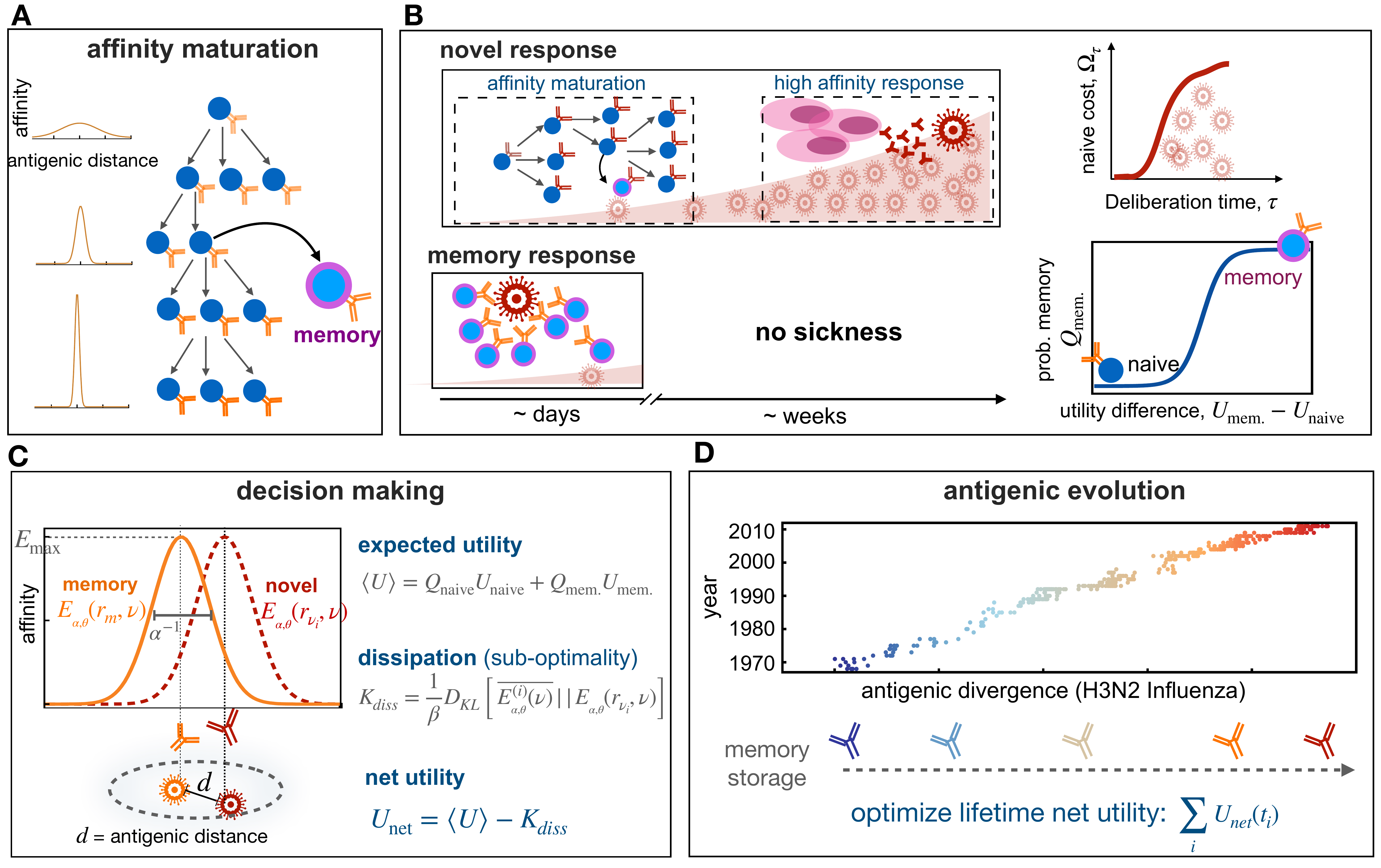}
\caption{\label{fig:Fig1}
{\bf  Immune memory or na\"ive response upon infection.} {\bf(A)} Schematic shows affinity maturation in germinal centers(right), where B-cell receptors acquire mutations and undergo selection, resulting in an increase in their affinity to an antigen (from light to dark receptors), indicated by the sharpening of receptors' affinity profiles (on left).  {\bf (B)} Upon  infection, the immune system can initiate  a novel response (top) or a memory response (bottom). A novel B-cell response could involve affinity maturation to generate memory or high-affinity  plasma cells (pink) that can secrete antibodies to battle the pathogen. A novel response can take 1-2 weeks, during which pathogen can replicate within a host and  a patient can show symptoms from the disease (top, left). During this time, the proliferation of pathogens within a host incurs a cost associated with a naive response  $\Omega_\tau$,  which is a monotonic function of the deliberation time $\tau$ (top, right). If the host carries memory from a previous infection or vaccination (bottom),  the immune system can robustly and rapidly activate a memory response to battle the infection. The probability to mount such memory response  $Q_{\text{mem.}}$ depends non-linearly on the relative utilities of memory versus na\"ive responses against a given infection  $\Delta U = U_\text{mem.}- U_\text{naive}$ (bottom, right).
 {\bf (C)}  Affinity profile  $E_{\alpha,\theta} (r_m,\nu) \sim \alpha\,\exp[- (\alpha d)^\theta]$ of a memory receptor $r_m$ is shown in orange as a function of the distance {$d=\Vert \nu^*_r-\nu \Vert$} in the antigenic shape space, between the receptor's cognate antigen $\nu^*_r$ (orange) and an evolved novel target  $\nu_i$ (red). The affinity of a receptor decays with increasing  distance between targets and its cognate antigen. The antigenic range over which a receptor is reactive inversely depends on its specificity $\alpha$. The shape of the binding profile is tuned by the factor $\theta$, here shown for $\theta=2$. The expected  binding profile $\overline {E^{(i)}_{\alpha,\theta}(\nu) }$ and the expected utility  $\langle U\rangle$ for an immune response are weighted averages of these quantities over memory and na\"ive responses. The Kullback-Leibler distance between the expected profile $\overline {E^{(i)}_{\alpha,\theta}(\nu) }$ and the profile centered around the infecting antigen $E_{\alpha,\theta} (r_{\nu_i},\nu)$, in units of the deliberation factor $\beta$, defines the sub-optimality of a response, i.e., dissipation $K_\text{diss}$  (eq.~\ref{eq:Dissipaion}). The net utility $U_\text{net}$ measures the goodness of a decision to mount a memory vs. naive response against an infection (eq.~\ref{eq:Unet}). {\bf (D)} Antigenic evolution of the H3N2 influenza virus is shown over 40 years along its first (most variable) antigenic dimension  (data from~\cite{Bedford:2014bf}). The decision of an immune system to utilize memory or to mount a novel response (B,C) is determined by the specificity $\alpha$ of receptors and the deliberation factor $\beta$. We characterize the optimal immune strategies ($\alpha^*,\beta^*$) by maximizing the total net utility of immune responses against pathogens with different antigenic divergences,  experienced over the lifetime of an organisms (eq.~\ref{eq:Unet_opt}).}
\end{figure*}

Beside receptor diversity,  immune cells also differentiate and specialize to take on different roles, including plasma B-cells, which are antibody factories, effector T-cells, which can actively battle infections, or memory cells.  Memory responses are highly efficient since memory cells can be reactivated faster than na\"ive cells and can mount a more robust response to an infection~\cite{McHeyzerWilliams:2000dd,Tangye:2003tf,Tangye:2004br,Moens:2016kc}. Memory generation is a form of cell fate decision in the immune system, which can occur at different stages of an immune response. In B-cells, activated na\"ive cells  can differentiate into  antibody-secreting long-lived plasma cells, a T-cell independent un-hypermutated memory cells, or they can initiate a germinal center~\cite{Goodnow:2010eo}. B-cells that enter germinal centers differentiate during affinity maturation into high-affinity plasma cells or T-cell dependent long-lived memory cells that circulate in the blood for antigen surveillance; see schematic~Fig.~\ref{fig:Fig1}. 

The basis for differentiation of B-cells into memory, especially during affinity maturation, is among the least understood in cell fate decision-making in the immune system~\cite{Goodnow:2010eo}. A long-standing view was that memory is continuously produced during affinity maturation~\cite{Blink:2005cg}. Memory receptors often have lower affinity compared to plasma cells~\cite{Smith:1997eva}, and therefore, if {memory B-cells} were to be generated continuously it should be able to proliferate without strong affinity dependent selection~\cite{Goodnow:2010eo,Victora:2012gx}. However, recent experiments indicate that memory differentiation is highly regulated~\cite{Paus:2006jg,Weisel:2016ep,Shinnakasu:2016ei,Recaldin:2016ir,Shinnakasu:2017ct,Viant:2020jt}, reflecting a temporal switch in germinal centers that preferentially produces memory at early stages and plasma at later stages of affinity maturation~\cite{Weisel:2016ep}. This active regulation introduces an affinity-dependent cell fate decision, leading to a preferential selection of low-affinity cells to the memory compartment. Low-affinity memory may be at a disadvantage in mounting a protective immune response since immune-pathogen recognition is largely determined by the binding affinity between an immune receptor and antigenic epitopes.  {On the other hand, immune-pathogen recognition is cross-reactive, which would allow memory receptors to recognize slightly evolved forms of the antigen, in response to which they were originally generated. }

We propose that the program for differentiation of immune cells to memory should be viewed in light of the immune system's coevolution with pathogens. We have developed a theoretical framework that incorporates the kinetics and energetics of memory responses as ingredients of memory strategy, which we seek to optimize under various evolutionary scenarios. We propose that the hard-wired affinity-dependent regulatory measures for  memory differentiation could be understood as a way to  optimize the long-term utility of immune memory against evolving pathogens. Individuals encounter many distinct pathogens with varying evolutionary rates, ranging from relatively conserved pathogens like chickenpox to rapidly evolving viruses like influenza. To battle such a spectrum of evolving pathogens,  we propose that an optimal immune system should store a combination of low-affinity memory with high cross-reactivity to counter evolving pathogens, and high-affinity and specific memory to counter the relatively conserved pathogens--- a strategy consistent with B-cell memory, which often involves storage of both cross-reactive IgM and high-affinity IgG receptors~\cite{Shlomchik:2020ab,McHeyzerWilliams:2018fv}.  Lastly, we study the impact of organisms' life expectancy on their evolved memory strategies and predict that cross-reactive memory should dominate the immune response in short-lived organisms that encounter only a few pathogens.

Previous work on  theoretical modeling of cellular differentiation together with experiments has been instrumental in understanding  immune memory generation; e.g. see reviewed work in~\cite{Perelson:1997dl,AltanBonnet:2020hk}. For example, mechanistic models  have indicated the importance of signal integration at the cellular level~\cite{Laffleur:2014ks}  and the relevance of stochastic effects at the population level~\cite{Hawkins:2007co}, to explain heterogeneous cell fate decisions for the generation of memory. Our statistical framework aims  to characterize high-level features for an optimal memory strategy, without relying on mechanistic details of  the underlying process, some of which are at least partially unknown~\cite{Bialek:2012ue,NourMohammad:2013in}. In the case of the immune system, statistical models  have provided an intuition for how  an  immune repertoire should be organized to optimally counter diverse pathogens~\cite{Perelson:1979ty,Mayer:2015ce,Mayer:2016jm,Mayer:2019is,Bradde:2020kb}. In a similar fashion, optimal memory strategies identified by our model provide a baseline to gauge the performance of real immune systems in storing and utilizing memory.

\section{Model} 
The efficacy of an immune response to a pathogen is determined by two key factors: (i) the affinity of immune-pathogen recognition (i.e., energetics), and (ii) the speed of response (i.e., kinetics) to neutralize an infection. 

Recognition of a pathogen (or its antigenic epitope) $\nu$ by an immune receptor $r$ is mediated by the affinity of the molecular interactions $E(r,\nu)$ between them.  We describe cross-reactive immune-pathogen recognition in an immune {\em shape space}~\cite{Perelson:1979ty}, where receptors located near each other in shape space can recognize similar antigens, and in the complementary space, antigens that are close to each other can be recognized by the same immune receptor (Fig.~\ref{fig:Fig1}). We express the binding affinity between a receptor $r$ and an arbitrary target antigen $\nu$ in terms of the antigenic distance $d_{r}(\nu) = \Vert \nu-\nu^*_r\Vert$ between the receptor's cognate antigen $\nu^*_r$ and the target $\nu$: $E(r,\nu ) \equiv E(d_{r}(\nu))$. 

Physico-chemical constraints in protein structures can introduce a tradeoff between immune receptors' affinity and {cross-reactivity}. Although we lack a systematic understanding of these structural constraints, affinity-specificity tradeoffs have been reported repeatedly for B-cells and antibodies~\cite{Wedemayer:1997co,frank:2002,Li:2003,Wu:2017ek,Mishral:2018,Fernandez:2020}. Specifically, while affinity maturation can significantly increase the binding affinity of a B-cell receptor, it also makes the receptor more rigid and specific to its cognate antigen~\cite{Wedemayer:1997co,Li:2003,Mishral:2018,Fernandez:2020}.  Broadly neutralizing antibodies (bNAbs) appear to be an exception to this rule since they have high potency and can react to a broad range of viral strains. However, it should be noted that bNAbs often react to vulnerable regions of a virus where escape mutations are very deleterious, including the CD4 binding site of HIV or the stem proteins in influenza~\cite{Mascola:2013vv,Lee:2015ij}.  In other words, the majority of bNAbs are not cross-reactive per se, but they are exceptionally successful in targeting conserved epitopes in otherwise diverse viral strains.

To qualitatively capture this affinity-specificity tradeoff, we use a simple functional form: We assume that the binding affinity of a receptor $r$ to an antigen $\nu$ depends on the antigenic distance $d_{r}(\nu)$ through a kernel with a specificity factor $\alpha$ and a shape factor $\theta$ such that, $E(r,\nu)\equiv E_{\alpha,\theta}(d_r(\nu)) \sim{\alpha} \exp[- \left( \alpha d_r(\nu)\right)^\theta] $, with $\theta \geq 0$. This affinity function defines a receptor's binding profile over the space of antigens.  As  specificity $\alpha$ increases (or cross-reactivity $1/\alpha$ decays), the binding affinity profile sharpens and binding becomes more restrictive to antigens closer to the receptor's cognate antigen (Fig.~\ref{fig:Fig1}). Moreover, the absolute strength of binding to the cognate antigen (i.e., a receptor's maximum affinity) increases with specificity $\alpha$, resulting in a tradeoff between affinity and cross-reactivity. The parameter $\theta$ tunes the shape of the receptor's binding profile  $E_{\alpha,\theta}(d_r(\nu))$, resulting in a flat function (i.e., no tradeoff) for $\theta=0$, a double-sided exponential function for $\theta=1$, a Gaussian (bell-curve) function for $\theta =2$, and top-hat functions for $\theta \gg 2$; see Materials and methods.\\ 

Upon encountering a pathogen, the adaptive immune system mounts a response by activating the na\"ive repertoire (i.e., a novel response) and/or by triggering previously stored immune receptors in the memory compartment. A memory receptor often shows a reduced affinity in interacting with an evolved form of the pathogen. Nonetheless, memory plays a central role in protecting against re-infections since even a suboptimal memory can be kinetically more efficient than a na\"ive response, both in B-cells~\cite{Tangye:2004br} and T-cells~\cite{Whitmire:2008fl,Martin:2012hf}. Specifically, following an infection, memory B-cells initiate cell division  about $1-2$ days earlier, and they  are recruited to proliferate in $2-3$ times larger numbers compared to the na\"ive population~\cite{Tangye:2003tf,Tangye:2004br,BlanchardRohner:2009dh}. Once recruited, however, memory and na\"ive cells have approximately a similar doubling time of about  $t_{1/2}\approx  0.5-2$ days~\cite{Tangye:2003tf,Macallan:2005hn}. Taken together, we can define an effective deliberation time  $\tau\approx 1.5-5$ days for the na\"ive population to reach an activity level (i.e., a clone size) comparable to the memory; see materials and methods and Fig.~\ref{fig:Fig1}.

The decision to mount a na\"ive or a memory response depends on the energetics and the kinetics of the immune machinery, including the cross-reactivity of memory to recognize evolved pathogens and the deliberation time to mount a na\"ive response upon infection --- we refer to these choices as {\em memory strategies}. We expect that the biochemical machinery involved in making this decision upon an infection has been fine-tuned and selected over evolutionary time scales in order to utilize immune memory and mount an effective response against recurring pathogens. The theory of decision-making~\cite{vonNeumann:1944tc,Ortega:2013eu} enables us to characterize the response of the immune system as a rational  decision-maker that chooses between two possible actions $a \in \{\text{na\"ive, memory}\}$ each contributing a {\em utility} $ U_a$ (Methods). Specifically, the action of a rational decision-maker should follow an optimal distribution $Q_a$, which maximizes the expected utility while satisfying the constraints in processing new information e.g. due to prior preferences~\cite{vonNeumann:1944tc,Ortega:2013eu}. We assume that the immune system has no intrinsic prior for mounting a na\"ive or a memory response against a given pathogen. In this case, the utility $U_a$ of an action (memory vs. na\"ive) determines the type of response, and rational decisions follow a maximum entropy distribution $Q_a \sim \exp[\beta U_a]$~\cite{Jaynes:1957fy}, where $\beta$ is the efficacy of information processing (see Methods). As $\beta$ increases, a rational decision-maker  more readily chooses the action with the highest utility. The expected utility of the immune response to an infection is equal to the sum of the utilities of a na\"ive and a memory response, weighted by their respective probabilities: $\left \langle U \right \rangle = U_\text{mem} \,Q_\text{mem.}+ U_\text{na\"ive}\, Q_\text{na\"ive}$. If memory is effective, the utility difference between  mounting a memory or a na\"ive response is determined by the affinity of the interaction between the responding memory receptor $r_m$ and the infecting antigen $\nu$: $U_\text{mem} -U_\text{na\"ive}=E_{\alpha,\theta}(r_m,\nu)$; see Fig.~\ref{fig:Fig1} and Methods for details. 

The time lag (deliberation) between memory and naive response also plays a key role in the decision-making process. On the one hand, if memory is inefficient, long deliberations would allow pathogens to proliferate, incurring a larger cost $\Omega_\tau$ to a host prior to activation of a novel response;  this cost can be interpreted as the negative utility of na\"ive response $U_\text{na\"ive} \equiv -\Omega_\tau$. On the other hand, a long deliberation would allow the immune system to exploit the utility of a usable memory (i.e., process information), even if the available memory  has only a slight advantage over a responsive na\"ive receptor (see Methods). Indeed, for a responsive memory, the information processing factor $\beta$ is equal to accumulated pathogenic load $\Gamma_\tau$ during the deliberation period $\tau$, and thus, we refer to $\beta$ as the {\em deliberation factor.} 

The expected binding profile of stored memory $\overline {E^{(i)}_{\alpha,\theta}(\nu)}$ after $i^{th}$ round of re-infection with an antigen $\nu_i$ can be characterized as the superposition of the binding profiles following a memory or a naive response, weighted  by the respective probability of each of these events (Fig.~\ref{fig:Fig1} and Methods). Since mounting a sub-optimal memory against evolved variants of a reinfecting pathogen can still be kinetically favorable, the expected profile can deviate from the optimal profile of the cognate receptor centered around the infecting pathogen $E_{\alpha,\theta}(r_{\nu_i},\nu)$ (Fig.~\ref{fig:Fig1}). This tradeoff between the kinetics and the energetics of immune response results in a {\em non-equilibrium decision-making}~\cite{GrauMoya:2018gp} by the immune system (Methods).
In analogy to non-equilibrium thermodynamics, we express this deviation as a dissipative cost of memory response $K_{\diss} (t_i;\alpha,\theta)$ at the $i^{th}$ round of re-infection (time point $t_i$), which we quantify by the Kullback-Leibler distance between the expected and the optimal binding profiles {$D_{KL}\left (\,\overline{ E^{(i)}_{\alpha,\theta}(\nu)} || E_{\alpha,\theta}(r_{\nu_i},\nu)\right)$, in units of the deliberation factor $\beta$ (Fig.~\ref{fig:Fig1})},
\EQA
\label{eq:Dissipaion}
\nonumber K_{\diss} (t_i) &=& \frac{1}{\beta} D_{KL}\left (\,\overline{ E^{(i)}_{\alpha,\theta}(\nu)} || E_{\alpha,\theta}(r_{\nu_i},\nu)\right) \\
\nonumber&=&\frac{1}{\beta}\sum_{\text{antigens: }\nu} \overline{ E^{(i)}_{\alpha,\theta}(\nu) }\log\left[\frac{\overline{ E^{(i)}_{\alpha,\theta}(\nu)} }{E_{\alpha,\theta}(r_{\nu_i},\nu)}\right].\\
\EEA

 \begin{figure}[t!]
\begin{center}\includegraphics[]{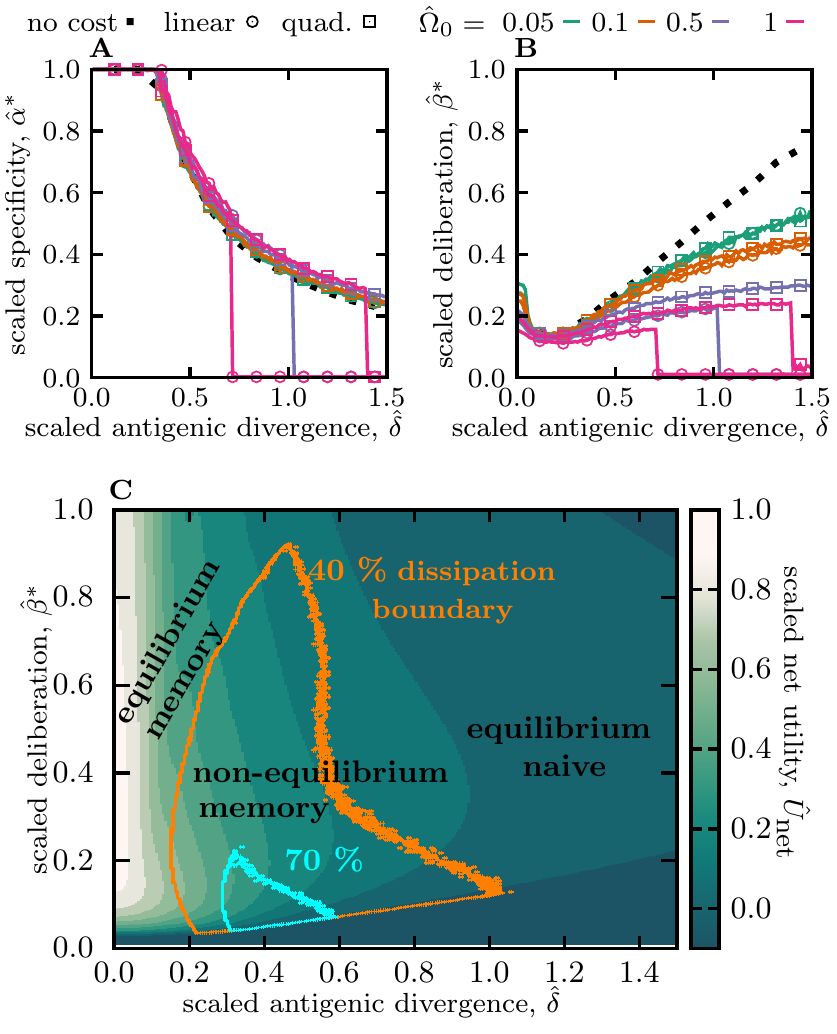}
\end{center}
\caption{\label{fig:Fig2} {\bf Optimal memory strategies against evolving pathogens.} {\bf (A)} and {\bf (B)} show the optimal  specificity  $\hat \alpha^* \equiv \alpha^*/\alpha_\text{max}$ and  deliberation factor $\hat \beta^*\equiv\beta^* /\beta_\text{max}$, scaled by their respective upper bounds,  as a function of the antigenic divergence per infection, scaled by the cross-reactive range (or inverse of maximum specificity)  $\hat\delta \equiv \delta /(\alpha_\text{max}^{-1})$.  Colors / markers indicate different na\"ive cost functions for deliberation, including no-cost $\hat \Omega\equiv\Omega/E_\text{max} =0$, linear cost $\hat \Omega  =\hat \Omega_0 \hat \beta$, and quadratic cost $\hat \Omega= \hat \Omega_0 \hat \beta^2$, with varying amplitudes $\Omega_0$. {\bf (C)} The heat map shows the  expected rescaled net utility {$\hat U_\text{net} =U_\text{net}/E_\text{max} $} (eq.~\ref{eq:Unet}) per round of infection for an immune system with an optimal specificity $\hat\alpha^*$, as a function of rescaled antigenic divergence $\hat \delta$ and deliberation factor $\hat \beta$. Rescaling by $E_\text{max}$ sets the magnitude of net utility to one,  for a response to conserved antigens (with $\hat \delta =0$) and in the limit of zero  deliberation cost $\Omega\to 0$. Boundaries indicate different levels of dissipation, with orange and blue encompassing regions of $\geq 40\%$ and $\geq 70\%$ of the  maximum dissipation $K_\text{max}$, respectively. The three modes of immune response are indicate based on the magnitude of dissipation and  net utility in each reagion: (i) equilibrium memory, (ii) non-equilibrium memory and (iii) equilibrium na\"ive. Simulation parameters,  (A-C): $\alpha_\text{max}=4$, $\beta_\text{max}=10$, and $\theta=2$, (C): linear deliberation cost function $\hat \Omega = \hat \Omega_0 \hat \beta$ with $\hat \Omega_0 = 0.1$. Results for other shape parameters $\theta$ and specificity thresholds $\alpha_\text{max}$ are shown in Figs.~\ref{fig:2sf2},~\ref{fig:2sf3}, respectively.}
\end{figure}
An optimal memory strategy should be chosen such that it maximizes the expected utility of the immune response $\langle U\rangle$, while minimizing the dissipation cost due to the non-equilibrium response $K_{\diss}$, over the lifetime of an organism. To infer an optimal strategy, we introduce net utility that accounts for the tradeoff between the expected utility and dissipation at a given round of infection at time point $t_i$,
\begin{equation}
\label{eq:Unet}
 \U_\text{net} (t_i)=\left \langle U(t_i) \right\rangle - K_{\diss}(t_i)
\end{equation}
We infer the optimal memory protocol (i.e., the optimal memory specificity $\alpha^*$ and deliberation factor $\beta^*$) by maximizing the total net utility of memory responses throughout the lifetime of an organism (Fig.~\ref{fig:Fig1}), 
\begin{equation}
 (\alpha^*,\beta^*)= \underset{\alpha,\beta}{\text{argmax} }\sum_{i:\text{ infections}} \U_\text{net} (t_i).
\label{eq:Unet_opt}
\end{equation}

\section{Results}
\noindent{{\bf Efficient immune memory balances specificity and speed.}}  The extent of cross-reactivity and deliberation needed for the memory to react to pathogens should be set by the amount of pathogenic evolution and more specifically, the antigenic divergence {\small $\hat\delta\equiv \sqrt{\langle \Vert\nu_i - \nu_{i-1}\Vert^2\rangle}$} that a pathogen traces between two infections. An example of such antigenic divergence is shown in Fig.~\ref{fig:Fig1}D for 40 years of H3N2 Influenza evolution along it first  (most variable) evolutionary dimension~\cite{Bedford:2014bf}. We set to find an optimal immune protocol (i.e., specificity $\alpha^*$ and deliberation $\beta^*$) by maximizing the net utility $ \U_\text{net}$ of an immune system (eq.~\ref{eq:Unet_opt}) that is trained to counter pathogens with a given antigenic divergence {$\hat\delta$}; see Fig.~\ref{fig:Fig1}D and Materials and methods for details on the optimization procedure. 

To battle slowly evolving pathogens  ($\hat\delta \leq 20\%$) an optimal immune system stores highly specific memory receptors, with a specificity that approaches the upper bound $ \alpha_\text{max} $; see  Figs.~\ref{fig:Fig2}A and ~\ref{fig:2sf2},~\ref{fig:2sf3}. Importantly, the dependency of optimal specificity on antigenic divergence is insensitive to the cost of deliberation $\Omega$ prior to mounting a na\"ive response (Fig.~\ref{fig:Fig2}A), the shape factor $\theta$ for the specificity profile (Fig.~\ref{fig:2sf2}), and the specificity threshold $\alpha_\text{max}$ (Fig.~\ref{fig:2sf3}). For relatively conserved pathogens ($\hat \delta \simeq 0$), the highly specific memory (with $ \hat \alpha^*\equiv \alpha^*/\alpha_\text{max} \simeq 1$) stored from a previous infection still has high affinity and remains centered and close to the reinfecting pathogens. Therefore, the immune system  maintains a moderate level of deliberation to exploit this efficient memory during infections. However, as antigenic divergence grows, specific memory becomes less effective against future infections and therefore, the immune system reduces the deliberation factor  to allow a timely  novel response,  once memory becomes inefficient (Figs.~\ref{fig:Fig2}B,~\ref{fig:2sf2},~\ref{fig:2sf3}). The magnitude of deliberation decays as the  cost of deliberation $\Omega$ increases but its overall  dependency on antigenic divergence remains  comparable for different cost functions (shown in Fig.~\ref{fig:Fig2}B for zero cost, and cost functions that grow  linearly and quadratically  with  deliberation factor $\beta$).  Overall, the net utility of the stored memory in response to slowly evolving pathogens is  high (Figs.~\ref{fig:Fig2}C,~\ref{fig:2sf1},~\ref{fig:2sf2},~\ref{fig:2sf3}), while  its dissipation  remains  small $K_\text{\diss} \simeq 0$  (Figs.~\ref{fig:Fig2}C,~\ref{fig:2sf1},~\ref{fig:2sf2},~\ref{fig:2sf3}). Therefore, in analogy to thermodynamics,  we term this immune strategy with low dissipation as   {\em equilibrium memory response}; Fig.~\ref{fig:Fig2}C. 

To battle moderately evolving pathogens (with {$\hat \delta \simeq 20\%-60\%$)}, an optimal immune system stores cross-reactive memory (i.e., with a lower specificity $\hat \alpha$)  that can recognize moderately evolved form of the primary antigen (Figs.~\ref{fig:Fig2}A,~\ref{fig:2sf2},~\ref{fig:2sf3}).  However, cross-reactive receptors tend to have lower affinities~\cite{Wedemayer:1997co,frank:2002}, which could lead to deficient responses  against antigens. Importantly, activation of energetically sub-optimal yet cross-reactive memory could be detrimental as it may hinder a stronger novel response without providing  protective immunity to the host--- a deficiency  known as the original  antigenic sin~\cite{FrancisJR:1960dc,Vatti:2017js}. An optimal immune system can mitigate this problem by using  kinetic optimization to tune the deliberation factor $\beta$ in order to avoid an elongated memory engagement prior to a na\"ive response. This optimization results in a smaller deliberation factor $\beta$ (i.e., a faster na\"ive response) compared to the scenario with slowly evolving pathogens, yet a long enough deliberation to allow the energetically suboptimal memory to react to an infection, whenever feasible (Figs.~\ref{fig:Fig2}B,~\ref{fig:2sf2},~\ref{fig:2sf3}). With this kinetic optimization, the immune system can utilize  cross-reactive memories through multiple rounds of infection (Fig.~\ref{fig:2sf1}C), yet with a declining efficiency and net utility as  pathogens evolve away from the primary infection (Figs.~\ref{fig:Fig2}C,~\ref{fig:2sf1},~\ref{fig:2sf2},~\ref{fig:2sf3}). The prominent memory response to moderately evolving pathogens is dissipative with $K_\text{\diss}\gg0$ (Figs.~\ref{fig:Fig2}C,~\ref{fig:2sf1},~\ref{fig:2sf2},~\ref{fig:2sf3}), and in analogy with thermodynamics, we term this dissipative immune strategy as  {\em non-equilibrium memory response}; Fig.~\ref{fig:Fig2}C. 

For extremely rapidly evolving pathogens ({$\hat \delta> 60\%$}), the  immune system would not be able to store an efficient memory to battle future encounters, and hence, each infection would trigger a novel na\"ive response --- the reduced net utility of memory and the  decay of memory usage in this regime are shown in Figs.~\ref{fig:Fig2}C,~\ref{fig:2sf1},~\ref{fig:2sf2},~\ref{fig:2sf3}, respectively. Without a protective memory, a novel response is triggered to counter each infection and it maturates specifically around the infecting pathogen, resulting in a   non-dissipative na\"ive-dominated immune response with $K_\text{\diss} \simeq 0$, which we term {\em equilibrium na\"ive response}; Fig.~\ref{fig:Fig2}C. 

It should  be noted that when the cost of deliberation $\Omega$ is very high, utilizing memory against pathogens with relatively high evolutionary rates  becomes highly unfavorable. In this extreme case, the immune system switches into a state where it invariably mounts a  novel response upon an infection (Fig.~\ref{fig:2sf1}C), and it assures that memory is not utilized by setting the parameters for specificity $\alpha$ and deliberation $\beta$ to zero (Fig.~\ref{fig:Fig2}A, B).

Our analyses in Fig.~\ref{fig:Fig2} indicate that a rational decision to become a memory or a plasma cell during an immune response should depend on the affinity of a cell's receptors and it should not be a stochastic choice with a constant rate throughout affinity maturation.  Indeed, cell fate decision for B-cells during affinity maturation is highly regulated and dependent on receptors' affinity~\cite{GoodJacobson:2010bf,Kometani:2013et,Shinnakasu:2016ei,Weisel:2016ep,Shinnakasu:2017ct,Shlomchik:2019kh}. Recent experiments have demonstrated that memory generation is highly correlated with the activity of the transcription factor {\em Bach2} whose expression level is negatively regulated with the abundance of helper CD4$^+$ T-cells~\cite{Kometani:2013et,Shinnakasu:2016ei,Shinnakasu:2017ct}. As the affinity of B-cell receptors increases during affinity maturation,  more  CD4$^+$ T-cells are recruited to germinal centers, resulting in suppression of {\em Bach2} and a hence, a decline in production of memory cells~\cite{Kometani:2013et,Shinnakasu:2016ei,Shinnakasu:2017ct}.  In other words, our adaptive immune system has encoded a negative feedback mechanism to store memory with intermediate affinity and cross-reactivity to suppress the production of highly specific memory, which is likely to be impotent against evolved pathogens in future infections. \\

\noindent{{\bf A mixture  memory strategy is necessary to  counter  pathogens with  a broad range of evolutionary rates.}} The decision to trigger an equilibrium or a non-equilibrium  memory response depends on the extent of antigenic divergence that an immune system is trained to cope with (Figs.~\ref{fig:Fig2},~\ref{fig:2sf1},~\ref{fig:2sf2},~\ref{fig:2sf3}). Equilibrium memory is highly effective (i.e., it has high net utility) against relatively conserved pathogens, however, it fails to counter evolving pathogens (Fig.~\ref{fig:Fig2}C). On the other hand, cross-reactive non-equilibrium memory is more versatile and can counter a broader range of evolved pathogens  but at a cost of  reduced   net utility in immune response; Figs.~\ref{fig:Fig2}C,~\ref{fig:2sf1},~\ref{fig:2sf2},~\ref{fig:2sf3}. 
\begin{figure}[t!]
\begin{center}
\includegraphics[]{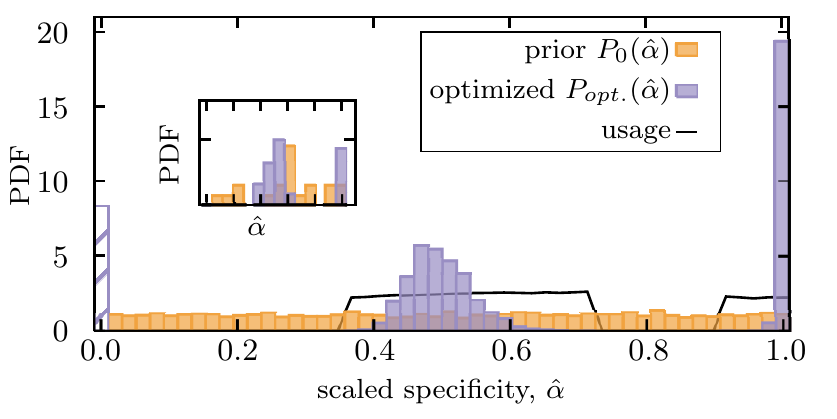}
\end{center}
\caption{\label{fig:Fig3}
{\bf Mixed memory strategy against  {a mixture of  pathogens with a broad range of evolutionary rates.}} 
Distribution of  scaled optimized specificities $\hat\alpha^*$ for functional memory (purple) is shown for an immune system with a fixed deliberation factor $\hat \beta = 0.2$. A mixture strategy with a bimodal distribution of specificities  $P(\hat \alpha)$ is established to counter  pathogens with a broad range of antigenic divergences. The dashed bar indicates stored memory with specificity $\alpha=0$, which is not further used in response to infections. The solid line indicates the probability $P_\text{usage}$ that a stored  memory with a given specificity is utilized in future infections (Methods). Optimization is done by maximizing the net utility of immune response averaged  over encounters with 1000 independently evolving antigens with (scaled) antigenic divergences drawn uniformly  from a range $ \hat \delta \in (0,1.6)$ (Methods). The distribution shows the ensemble statistics of functional memory accumulated from 200 independent optimizations, each starting from a flat prior for specificities (orange). The insert shows the optimized mixture strategy for one  optimization with 3000 steps.  Simulation parameters:  $\alpha_\text{max}=4$, $\beta_\text{max}=10$, and $\theta=2$.}\end{figure}

An optimal immune system should have memory strategies to counter pathogens with varying evolutionary rates, ranging from relatively conserved pathogens like chickenpox to rapidly evolving viruses like influenza. We use our optimization protocol to find such memory strategies that maximize the net utility of  an immune system that  encounters evolving pathogens with (scaled) antigenic divergences uniformly drawn from a  broad range of  $\hat\delta \in[0 \,\,1.6] $; see Materials and methods. This optimization results in a bimodal distribution of optimal specificity for functional memory receptors $P(\alpha)$, with separated peaks corresponding to equilibrium ($\hat \alpha \sim 1$) and non-equilibrium ($\hat \alpha \sim 0.5$) memory  (Figs.~\ref{fig:Fig3},~\ref{fig:3sf1}). This result suggests that specific and cross-reactive memory strategies are complementary modes of immune response that cannot substitute each other.  Moreover,  non-equilibrium  memory tends to be flexible and moderate values of  cross-reactivity $1/\hat\alpha$ can counter a range of  antigenic divergences, without a need for fine-tuning.  Therefore,  upon production of memory, an optimal immune system should harvest both specific equilibrium memory and cross-reactive non-equilibrium memory, as it does not have a priori knowledge about the evolutionary rate of the infecting pathogen.

Interestingly, the adaptive immune system  stores a mixture of   IgM and  class-switched IgG isotypes of B-cell memory that show different levels of specificity. IgM memory is an earlier product of affinity maturation with higher cross-reactivity and a lower affinity to antigens, reflecting a non-equilibrium memory that can counter evolving pathogens. On the other hand, memory from class-switched (e.g. IgG) isotype is produced during later stages of affinity maturation and is highly specific to the infecting pathogen, reflecting equilibrium memory that is effective against relatively conserved pathogens~\cite{Weisel:2016ep}. Storing a mixture of  IgM and class-switched IgG memory is consistent with our recipe for optimal immune strategies to counter  pathogens with a broad range of evolutionary rates.  \\

\noindent{{\bf Cross-reactive memory dominates immune response in organisms that encounter fewer pathogens over a shorter lifetime.}} So far, our analysis has focused on maximizing the net utility of immune response, assuming that organisms encounter many such infections throughout their lifetime. This optimization provides a recipe for optimal immune strategies in response to commonly infecting pathogens. However, the expected frequency of infections  is also an important factor that can inform immune strategies.  For example, imagine the extreme case that an immune system expects to encounter a pathogen at most only once during an organism's lifetime, e.g. in short-lived organisms.  In this case, there is no benefit in keeping a memory even to counter extremely conserved pathogens, for which memory would be otherwise very beneficial.

\begin{figure*}[t!]
 \includegraphics[width=\textwidth]{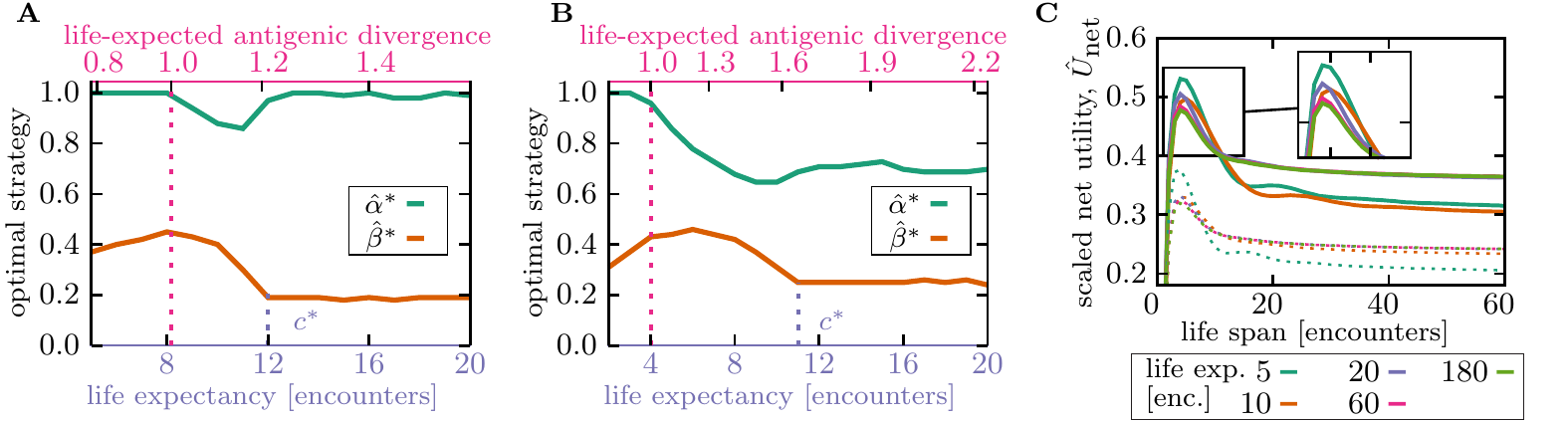}
\caption{\label{fig:Fig4}{\bf Life expectancy influences the specificity of optimal memory.} {\bf(A,B)} Memory strategies, i.e., optimal rescaled specificity $\hat \alpha^*$ (green) and deliberation factor $\hat\beta^*$ (orange) are shown as a function of the organism's life expectancy (bottom axis) and the corresponding expected antigenic divergence over the organism's life-time $\hat\delta \sqrt{\text{lifetime}}$ (top axis).  Antigenic divergence (per encounter) of the infecting pathogen is $\hat \delta =  0.35$ in (A) and  $\hat \delta =  0.5$ in (B). Memory is highly specific in organisms with very short lifetimes, during which  re-infections with evolved forms of a pathogen are unlikely (i.e., when life-expected antigenic divergence is smaller than 1, indicated by a dotted pink line). Memory becomes more cross-reactive with a smaller deliberation in organisms with (realistic) short lifetimes, up to a transition point  $c^*$ (indicated by dotted purple line), after which specificity increases again.  {\bf (C)} Scaled net utility $\hat U_\text{net}$  is  shown as a function of  organism's life span, whose immune strategies ($\hat\alpha^*$, $\hat\beta^*$)  are optimized for a specified life expectancy (colors as indicated in the legend).  Net utility for memory optimized against pathogens with antigenic divergence $\hat{\delta} = 0.35$ (panel A) and  $\hat{\delta}=0.5$ (panel B) are shown by full and dashed lines, respectively. Life span and life  expectancy are measured in units of the number of pathogenic encounters during lifetime. Simulation parameters:  linear deliberation cost function $\Omega= \Omega_0 \hat \beta$ with an amplitude $\hat \Omega_0 = 0.1$, $\alpha_{\text{max}}=4$, $\beta_{\text{max}}=10$, and $\theta=2$.}
\end{figure*}

To study the impact of infection frequency on immune strategies, we use our optimization procedure to maximize the net utility of immune response, while setting a bound on the number of infections throughout an organism's lifetime (see Methods). Organisms with an {\em unrealistically} very short lifetime (measured in units of the number of infections) experience only a few infections, and therefore, a small (cumulative) antigenic drift from the primary infection during their lifetime $\hat \delta \sqrt{\text{life time.}}\lesssim 1$.
In this case, it would be  sufficient for an optimal immune system to generate specific memory ($\hat\alpha\approx 1$), which can  mount an effective response  with only an intermediate deliberation ($\hat \beta\sim 0.4$) upon reinfection (Fig.~\ref{fig:Fig4}A-B), even for pathogens with a moderate evolutionary rate (Fig.~\ref{fig:Fig4}B). Organisms with  moderately short lifetime   experience evolutionary divergence of reinfecting antigens. In this regime, the immune system  stores cross-reactive memory (smaller $\hat\alpha$) and uses a larger deliberation factor $\hat \beta$ such that this lower-affinity and often off-centered memory can mount an effective response to evolved infections (Fig.~\ref{fig:Fig4}A-B). Since the organism is relatively short-lived, such cross-reactive memory could be sufficient throughout the whole lifetime  of the organism, without a need for renewal.

Organisms with long lifetimes, with pathogen encounters that surpassing the threshold $c^*$, expect higher re-infections with pathogens that are  highly diverged from the primary infection. In this case, an optimal immune strategy switches from storing and utilizing  cross-reactive memory to generating more specific memory receptors (Fig.~\ref{fig:Fig4}A). This specific memory would not hinder activation of  preventive novel responses against evolved pathogens (the problem known as original antigenic sin), resulting in continual renewal of memory during organisms' lifetime. In this regime, the deliberation factor also decreases to facilitate novel responses against antigens that are not readily recognized by memory (Fig.~\ref{fig:Fig4}A-B). The increase in memory specificity from short- to  long- lived organisms is more substantial for  immune strategies optimized to counter relatively conserved pathogens, i.e., the specific equilibrium memory (Figs.~\ref{fig:Fig2}C,~\ref{fig:Fig4}A), compared to the memory against evolving pathogens,  i.e., the  cross-reactive  non-equilibrium memory (Figs.~\ref{fig:Fig2}C,~\ref{fig:Fig4}B). The exact value of the transition threshold $c^*$ depends on the expected antigenic divergence $\delta$ during pathogenic evolution  and the details of the immune machinery, and specifically the cost of deliberation $\Omega(\tau)$ due to an elevated level of  pathogenic proliferation  prior to a novel response (Fig.~\ref{fig:4sf1}). However, the qualitative trend for  cross-reactivity as a function of the organism's lifetime remain consistent across a range of parameters.

The results in Fig.~\ref{fig:Fig4} predict that organisms with few pathogenic encounters or a shorter life-span should generate more cross-reactive and lower affinity (i.e., a na\"ive-type) memory receptors. Indeed, consistent with our prediction, analysis of immune repertoire data indicates that   sequence features of memory and  na\"ive B-cell receptors tend to be more similar to each other in mouse compared to humans that enjoy  a longer life expectancy~\cite{Sethna:2017bv}.  Nonetheless,  more comprehensive  data on  cross-species comparison of immune strategies is needed to test our predictions. 

With the increase in human life expectancy, a pressing question is how well our immune system could cope with a larger number of pathogenic challenges that we are now encountering throughout our lifetimes? Aging has many implications for our immune machinery and the history of infections throughout lifetime leaves a complex mark on immune memory that can have long-lasting consequences~\cite{Saule:2006ct}, which has also been studied through theoretical modeling~\cite{Mayer:2019is}. In our framework, we can study one aspect of this problem and ask how an immune strategy optimized to battle a given number of infections would perform if the organism were to live longer or equivalently, to encounter pathogens more frequently. Fig.~\ref{fig:Fig4}C shows that cross-reactive memory generated by an immune system optimized to counter few infections (short life expectancy) becomes highly inefficient (i.e., with a lower  net utility $\U_\text{net}$) as the number of encounters increases beyond the organism's expectation (long life span) --- an effect that may be in part responsible for the observed  decline in the efficacy of our adaptive immunity as we age.

\section{Discussion}
Memory is central to our adaptive immunity by providing a  robust and preventive response to reinfecting pathogens.  In the presence of continually evolving pathogens, immune memory is only beneficial if receptors can recognize evolved antigens by cross-reactivity. However, biophysical constraints can impose a trade-off between affinity and cross-reactivity of antibodies. Specifically, as receptors undergo affinity maturation, their structures become more rigid and less cross-reactive, while affinity increases~\cite{Wedemayer:1997co,frank:2002,Li:2003,Wu:2017ek,Mishral:2018,Fernandez:2020}.  Consistent with recent experiments~\cite{Weisel:2016ep,Shinnakasu:2016ei,Recaldin:2016ir,Shinnakasu:2017ct,Viant:2020jt}, we show that memory differentiation should be regulated to  preferentially produce  lower affinity receptors, which can  allow cross-reactive recognition of evolved pathogens. To overcome the resulting energetic impediment of these memory receptors, we infer that the immune system should  tune the kinetics of the immune response and allocate a longer deliberation time for memory to react before initiating a novel response --- a feature that is also  in accordance with observations~\cite{Tangye:2003tf,Tangye:2004br,BlanchardRohner:2009dh}. Co-optimizing kinetics and energetics of memory ensures an effective response against evolving  pathogens, throughout an organism's lifetime.

Optimal cross-reactive immune memory  provides a long-term advantage to  an organism, yet it may seem energetically sub-optimal over short time scales (Fig.~\ref{fig:Fig1}). One  important consequence of a  sub-optimal memory  response is known as  original  antigenic sin, where cross-reactive memory  from primary infections could interfere with and suppress a protective novel response~\cite{FrancisJR:1960dc,Vatti:2017js}. The viral exposure history and the original antigenic sin may have profound consequences on protective immunity against evolving viruses~\cite{Cobey:2017ku}. For example, the 2009 H1N1 pandemic triggered memory responses in individuals with childhood exposures  to seasonal H1N1~\cite{linderman_antibodies_2016,li_immune_2013,hensley_challenges_2014}, which in some led to a highly focused antibody response towards the conserved epitopes of H1N1. This focus was a problem when in 2013-2014 the pandemic H1N1 acquired mutations in those epitopes~\cite{linderman_antibodies_2016}, resulting in a disproportionate impact of infection on middle-aged individuals with pre-existing memory~\cite{petrie_antibodies_2016}. This recent example, among others, showcases  how immune history and antigenic sin can impact a population's immune response to the a rapidly evolving  virus like influenza.

Composition of the immune memory coupled with the exposure history of the host should be taken into account when designing new vaccines~\cite{Cobey:2017ku}. For example, current vaccine strategies against influenza use sera isolated from ferrets infected with the virus to measure the antigenic distance  of circulating strains against the previous years \cite{smith_mapping_2004}. However, these ferrets have no immune history for influenza and the  antibodies they produce may be distinct from  the immune response in the adult population with prior memory, resulting in incorrect measures of antigenic distances~\cite{hensley_challenges_2014}. This  problem has been recognized by the World Health Organization and there is now an effort to choose vaccine strains based on  human serology.

 The impact of  immune deficiency related to the original antigenic sin can even be more pronounced due to changes in an organism's life expectancy. Importantly, we show that immune strategies optimized to  benefit short-lived organisms produce highly cross-reactive memory (Fig.~\ref{fig:Fig4}). If an organism's life-expectancy increases, which is the case for humans,  it would be likely for  individuals to encounter evolved forms of a pathogen at antigenic distances larger than expected by their immune systems. In this case, cross-reactive memory, optimized for a shorter  lifetime, could still be activated but with  lower efficacy, which could suppress a protective novel response, consistent with original antigenic sin. It is therefore important to consider sub-optimality of immune strategies in the face of extensive elongation of the human  lifespan  as one of the plausible factors responsible for immune deficiencies brought by aging.

One characteristic of memory B-cells, which is currently missing from our model,  is their ability to seed secondary germinal centers and undergo further affinity maturation upon reinfection. Evolvability of memory B-cells can allow cross-reactive memory to further specialize against evolved pathogens, without a need to start a germinal center reaction from an un-mutated na\"ive receptor. Interestingly, different experiments suggest that the capacity of memory to re-diversify depends on various factors including the memory isotype (IgM vs. class-switch receptors), the type of antigenic target (viruses vs. others) and the extent of memory maturation~\cite{Shlomchik:2020ab,McHeyzerWilliams:2018fv}. Therefore, it is interesting to extend our model to study how evolvability of memory can influence its longterm utility to respond to evolving  pathogens, and especially viruses. 

{Evolvability of memory is also relevant for characterizing the dynamics of immune response to chronic viral infections like HIV. Analyses of immune repertoires  in HIV patients  over multiple years of infection have shown a rapid turnover and somatic evolution of B-cell clonal lineages to counter the evolution of the virus within hosts~\cite{Nourmohammad:2019ij}. It would be interesting to see  how the constant pressure from the evolving HIV on a host's immune system impacts the dynamics and  efficacy of immune memory over time. In addition, understanding the limits of memory re-diversification is instrumental in designing successive vaccination protocols with antigen cocktails to drive extensive affinity maturation of BCR lineages to elicit broadly neutralizing antibodies~\cite{Wang:2015em,Shaffer:2016ci,Stephenson:2020bw}--- an approach that is the  current hope for universal vaccines against rapidly evolving viruses like HIV. }

{Although mechanistically distinct from B-cells, T-cells also differentiate into effector and memory in response to infections. The T-cell response does not involve  affinity maturation by hypermutations. However,  competition among T-cells with varying receptor affinities acts as selection that leads to immuno-dominant responses by the high-affinity clones. Receptor affinity and the subsequent T-cell signaling determine the extent of clonal expansion and differentiation to an effector versus a memory T-cell population~\cite{Kim:2010im}. Although it is still unresolved as  how T-cell signaling determines cell fate decision,  the process is known to be highly regulated~\cite{Rutishauser:2009,Roychoudhuri:2016ni}. Notably, the transcription factor IRF4 selectively promotes expansion  and  differentiation of high-affinity cytotoxic T-cells into effectors. In contrast, low-affinity T-cells are lost or they could differentiate into early memory~\cite{Man:2013ni}. There is also accumulating evidence for the circulation of cross-reactive memory T-cells, which often result in protective immunity against evolving forms of a virus~\cite{greenbaum:2009,sette:2020}, but could also be detrimental  by suppressing novel and specific responses--- an effect similar to the original antigenic sin by B-cells~\cite{Selin:2004si}. Taken together, there are parallels between differentiation of T-cells and B-cells to memory, and it will be interesting to investigate the  advantages of storing cross-reactive (and plausibly low-affinity) T-cell memory as a strategy to counter evolving pathogens.} \\

\section*{Acknowledgements}
This work has been supported by the DFG grant (SFB1310) for Predictability in Evolution and the MPRG funding through the Max Planck Society. O.H.S also acknowledges funding from Georg-August University School of Science (GAUSS) and the Fulbright foundation.

\onecolumngrid
{\newpage{}
\setcounter{equation}{0}
\renewcommand{\theequation}{S\arabic{equation}}

\noindent {\Large\bf Supplementary Information}

\section*{\large i. Numerical optimization}
Numerical optimization is performed on ensembles  of immune systems that encounter evolving pathogens. Recognition of an evolved pathogen at the $i^{th}$ round of infection $\nu_i$ by a memory that was stored  in response to a primary infection $\nu_0$ ($0^{th}$ round) depends on the antigenic distance $d_i= \Vert\nu_i-\nu_{0}\Vert$.  We model pathogenic evolution as diffusion in the antigenic shape space. In this model, the expected antigenic distance between the primary infection $\nu_0$ and the evolved antigen $\nu_i$ can be characterized as, $ \langle d_i^2\rangle \equiv \langle \Vert\nu_i-\nu_{0}\Vert^2\rangle = \zeta^2(t_i - t_0) =i\,\delta^2$, where $\zeta$ is the diffusion coefficient (i.e., the evolutionary rate) and $\delta$ is the (averaged)  antigenic divergence per round of infection. Importantly, this relationship does not depend on the dimensionality of the antigenic shape space, which  in general, is difficult to characterize.  We simulate  pathogenic evolution relative to a primary infection by drawing the corresponding antigenic distance $d_i$ of the $i^{th}$ round of infection from a normal distribution with mean $\delta \sqrt{i}$ and standard deviation $0.05\delta  \sqrt{i}$. The width of this normal distribution characterizes the fluctuations in the mean divergence between infections and reflects how the evolutionary trajectory of a pathogen samples the multi-dimensional shape space surrounding the antigen from the primary infection. Nonetheless, our results  are insensitive to the exact choice of this width.

To characterize  optimal specificity $\alpha^*$ and deliberation factor $\beta^*$ (Figs.~\ref{fig:Fig2},~\ref{fig:Fig3},~\ref{fig:Fig4}), we simulate  ensembles of immune systems with different immune strategies ($\alpha, \beta$), chosen uniformly from the range  $\alpha \in [0, \alpha_{max}]$ and $\beta \in [0,\beta_{max}]$, with 500 increments in both parameters. Each immune system experiences  successive rounds of infection with an evolving pathogen with a given antigenic divergence $\delta$. During each encounter, the immune system chooses between utilizing an existing memory or initiating a novel response according to eq.~\ref{eq:recogProb}. The net utility of each encounter is calculated according to eq.~\ref{eq:Unet_SI}. We estimate the expected net utility per encounter over a lifetime of 60 total encounters and repeat this experiment across $10^5$ independent ensembles to find the optimal  immune strategies  $(\alpha^*, \beta^*)$ with the highest net utility. As shown in Fig.~\ref{fig:Fig4}, simulating up to 60 encounters is sufficient for the inference of optimal strategies in the asymptotic regime (i.e., a long lifetime).

To characterize optimal immune strategies against a mixture of pathogens with distinct levels of antigenic divergences, we define the mixture immune strategy by a set of specificities $\vec\alpha= \{\alpha_i\} =(\text{with, } i=1,\dots, N_m)$, where each $\alpha_i$  is a degree of specificity that a stored memory receptor can potentially have, and $N_m$ is the number of possible specificity strategies that an immune system can choose from.  The probability that an immune system with the mixture strategy $\vec\alpha$  recognizes a pathogen $\nu$ through a memory response  follows from an extension of eq.~\ref{eq:recogProb},
\EQA
\nonumber P^{(m)}_{\text{recog.}} (\vec\alpha, \nu) &=& 1-\prod_{\text{specificity: } {\alpha_i}} \left (1- P^{(m)}_{\text{recog.}} (r_m^{\alpha_i}, \nu) \right)\\
&=&1 -\prod_{\text{specificity: } {\alpha_i}}  e^{-E_{\theta} (r_m^{\alpha_i},\nu) \Gamma(\tau)} = 1- e^{-\sum_{\alpha_i} E_{\theta} (r_m^{\alpha_i},\nu) \Gamma(\tau)} \equiv 1- e^{- \tilde \beta  \, \overline{ E}_\theta(\nu)} 
\EEA
where $\overline{E}_\theta(\nu) = \frac{1}{N_m} \sum_{r_m^\alpha} E_{\theta} (r_m^{\alpha_i},\nu)$ is the expected affinity of  memory  (with distinct specificities)   against  antigen $\nu$ in an immune repertoire and $ \tilde \beta \equiv  N_m \beta$ is an effective deliberation factor for all choices of specificity. It should be noted that this effective deliberation factor $\tilde \beta$ is an extensive quantity with respect to the number of specificity strategies that an immune system can choose from, and therefore, is comparable across immune systems with different numbers of strategies.

We set out to characterize the mixture strategy as the probability $P_{\beta}(\alpha)$ based on which an immune system with a given effective deliberation factor $\tilde \beta$ should store a memory receptor with specificity $\alpha$, in order to optimally counter infecting pathogens with distinct antigenic divergences, drawn from a distribution $P(\delta)$. We start our optimization by defining a uniform mixture strategy, where the elements of the immune specificity vector $\vec\alpha =\{\alpha_i\}$ (of size $N_m=20$), are drawn uniformly from the range $[0,\alpha_\text{max}]$. Each optimization step aims to improve the specificity vector $\vec \alpha$ to maximize the net utility (per encounter) of the mixture  immune response $U_\text{net}(\vec\alpha^k)$ against 1000 independently evolving antigens whose (scaled) antigenic divergences are drawn uniformly from the range $\hat \delta = [0,\hat\delta_\text{max}]$.  We use stochastic simulations to estimate the net utility of the mixture strategy $U_\text{net}(\vec\alpha^k)$, whereby the relative affinity of  memory receptors (with varying specificities), $E_\theta(r_m^{\alpha_i},\nu)/ \overline{ E}_\theta(\nu)$,  determines the stochastic rate of their response to the infecting antigen $\nu$. The net utility (per encounter) of the immune response against each of the 1000 independently evolving antigens is estimated by averaging over a host's lifetime with 200 rounds of pathogenic encounters.     We update the mixture strategy over 3000 steps, using local gradient ascent by sampling 100 points  in the space of specificity vectors at each step to maximize  net utility, 
\EQA
\vec \alpha^{k+1}= \vec \alpha^k + \epsilon \nabla U_\text{net}(\vec\alpha^k)
\EEA
Here,  $k$ indicates the optimization step and $\epsilon=0.1$ is a hyper-parameter for gradient ascent.  We repeat the  optimization process starting from  200 independently drawn initial uniform mixture strategies $\vec \alpha^{\,0}$ to characterize the ensemble  of optimal memory strategies $P_{\beta}(\alpha)$ against pathogens with distinct  antigenic divergences drawn uniformly from a given range $\hat \delta = [0,\hat\delta_\text{max}]$, as shown in Fig.~\ref{fig:Fig3}. We also characterize the probability that a stored memory with a given specificity is  utilized against future infections (solid line in Fig.~\ref{fig:Fig3}). To do so, we  test the optimized ensemble of specificities $P_{\beta}(\alpha)$ against  5000 independent pathogens with antigenic divergences  drawn uniformly from the range $\hat \delta = [0,\hat\delta_\text{max}]$. We evaluate the usage of a memory with a given specificity $\alpha$ (solid line in Fig.~\ref{fig:Fig3}) as the conditional probability $P_\beta(\text{use } \alpha | \text{produce } \alpha)$ for using that memory given that it is produced (i.e., drawn from the distribution $P_{\beta}(\alpha)$).\\

{\noindent{\bf Code availability} All codes for simulations and numerical analysis can be found at: \\\href{https://github.com/StatPhysBio/ImmuneMemoryDM}{https://github.com/StatPhysBio/ImmuneMemoryDM}\\
}

\section*{\large ii. Model of evolutionary decision-making for adaptive immune response} 

\subsection*{ Kinetics of na\"ive  and memory immune response} 
Upon encountering a pathogen, the adaptive immune system mounts a response by activating the na\"ive repertoire (i.e., a novel response) and/or by triggering previously stored immune receptors in the memory compartment. A memory receptor often shows a reduced affinity in interacting with an evolved form of the pathogen. Nonetheless,  memory plays a central role in protecting against re-infections since even a suboptimal memory can be kinetically more efficient than a na\"ive response, both in B-cells~\cite{Tangye:2004br} and  T-cells~\cite{Whitmire:2008fl,Martin:2012hf}. First, memory cells are fast responders and initiate cell division about  $\tau_0\approx 1-2$ days before na\"ive cells~\cite{Tangye:2003tf,Tangye:2004br,BlanchardRohner:2009dh}. Second, the number of memory cells that are recruited to proliferate and differentiate to effector cells is $b\approx 2-3$ times larger than the number of na\"ive cells~\cite{Tangye:2003tf,Tangye:2004br}. Once recruited, however, memory and na\"ive cells have approximately a similar doubling time of about  $t_{1/2}\approx  0.5-2$ days~\cite{Tangye:2003tf,Macallan:2005hn}. Putting these kinetic factors together, we can define an effective deliberation time  $\tau$ for the na\"ive population to reach an activity level (i.e., a population size) comparable to the memory. Assuming an exponential growth during the early stages of memory and na\"ive proliferation,  the deliberation time can be estimated in terms of the kinetic factors by $\tau= \tau_0 + t_{1/2} \ln b/\ln 2$ and it is within a  range of  $\tau \approx 1.5-5$ days; see Fig.~\ref{fig:Fig1}.\\

\subsection*{Energetics of immune recognition}
 We assume that each immune receptor $r$ has a cognate antigen  $\nu_r^*$ against which it has the highest affinity. We express the binding affinity between a receptor $r$ and an arbitrary target antigen $\nu$ in terms of the antigenic distance $d_{r}(\nu) = \Vert \nu-\nu^*_r\Vert$ between the receptor's cognate antigen  $\nu^*_r$ and  the target $\nu$:  $E(r,\nu ) \equiv E(d_{r}(\nu))$. This distance-dependent binding affinity is measured with respect to the affinity of unspecific antigen-receptor interactions, sufficient to trigger a generic na\"ive response.

Physico-chemical constraints in protein structures can introduce a tradeoff between immune receptors' affinity and cross-reactivity (i.e., ability to equally react to multiple targets).  Prior to affinity maturation, the structure of na\"ive receptors is relatively flexible whereas hypermutations often reconfigure the active sites of a receptor and make them more specific so that they match their target antigens like a lock and key~\cite{Wedemayer:1997co,frank:2002}. As a result, the  IgM class of antibodies, which are the first line of defense in  B-cell response, often have low affinities, yet they are cross-reactive and can recognize mutated forms of the same epitope. On the other hand, the high-affinity IgG class of antibodies, which are the late outcomes of affinity maturation in germinal centers, have higher affinities but bind very specifically to their cognate antigen~\cite{frank:2002}.  Broadly neutralizing antibodies (bNAbs) are exceptions to this rule since they often have high potency and can react to a broad range of viral strains. However, bNAbs often react to vulnerable regions of a virus where escape mutations are very deleterious~\cite{Mascola:2013vv}.  In other words, the majority of bNAbs are not cross-reactive per se, but they are exceptionally successful in targeting conserved epitopes in otherwise diverse viral strains.  Nevertheless, an affinity-specificity tradeoff has been reported for a bNAb against the hemagglutinin epitope of  influenza~\cite{Wu:2017ek}. 

We use a simple functional form to qualitatively capture  the tradeoff between  cross-reactivity and affinity of antigen-receptor binding interactions: We assume that the binding affinity of a receptor $r$ to an antigen $\nu$ depends on the antigenic distance $d_{r}(\nu) = \Vert\nu-\nu_r^*\Vert$ through a kernel with a specificity factor $\alpha$ and a shape factor $\theta$ such that, $E(r,\nu)\equiv E_{\alpha,\theta}(d_r(\nu)) \sim{\alpha} \exp[- \left( \alpha \Vert\nu-\nu_r^*\Vert\right)^\theta] $, {with $\theta \geq 0$}. The width of this binding profile (i.e., the cross-reactivity) is set by the inverse of the specificity factor $ 1/\alpha$ (Fig.~\ref{fig:Fig1}), which decays as the height of the function (i.e., the maximum affinity) increases. The parameter $\theta$ tunes the shape of the receptor's binding profile  $ E_{\alpha,\theta}(d_r(\nu))$,  resulting in a flat function (i.e., no tradeoff) for $\theta=0$, a double-sided exponential function for $\theta=1$,  a Gaussian (bell-curve) function for $\theta =2$, and top-hat functions for  $\theta \gg 2$. Structural constraints and molecular features of protein receptors define a bound on the minimum cross-reactivity or  equivalently, a maximum specificity $\alpha_\text{max}$, achievable by a receptor.  Using this bound, we define rescaled  specificity  $\hat\alpha \equiv \alpha/\alpha_\text{max}$ to characterize the energetics of an immune response in a dimensionless form. \\

\subsection*{ Immune response to evolving pathogens}
Upon primary infection (i.e., an encounter with a novel pathogen)  na\"ive immune receptors with moderate affinity are activated to develop  a specific response through affinity maturation (Fig.~\ref{fig:Fig1}). Since the  na\"ive  repertoire is  diverse enough to contain receptors of moderate affinity against different antigens, we assume that the affinity  of  responsive na\"ive receptors, and hence, the strength of a primary immune response  to be approximately the same for all pathogens. This simplification becomes less accurate as the immune system ages  and the supply of effective receptors become more scarce.  

Following a na\"ive response to a primary infection and the subsequent affinity maturation, the immune system stores memory cells with an enhanced affinity to use them against future infections~\cite{Janeway:H7fnIHBf}; see Fig.~\ref{fig:Fig1}.  Therefore, the cognate antigen $\nu^*_{r_m}$ for a given memory receptor $r_m$ is an epitope derived from the primary infection that  led  to the formation of  memory, which we denote by $\nu_0$ with a subscript that indicates  round of infection. Thus, the binding profile  $E_{\alpha,\theta} (r_m,\nu)$ of the memory receptor $r_m$  is  peaked around the primary antigenic epitope $\nu^*_{r_m}=\nu_0$ (Fig.~\ref{fig:Fig1}). As pathogens evolve  globally to escape  the immune challenge, drugs, or vaccination, they drift away from the primary antigen  in antigenic space.    We model this antigenic shift as  a diffusion in shape space whereby a reinfecting pathogen at the $i^{th}$ round of infection $\nu_{i}$ is {\em on average}  at a distance $ \delta = \sqrt{ \langle \Vert\nu_{i} -\nu_{i-1}\Vert^2\rangle} $ from the previous infection $\nu_{i-1}$. This antigenic shift is proportional to the rate of pathogen evolution $\zeta_\nu $  and the average time between infections $\Delta t =t_i -t_{i-1}$, such that $\delta \propto \zeta_\nu \sqrt \Delta t$. A cross-reactive memory can  mount a response to an evolved  antigen, yet with a reduced affinity that decays with antigenic shift; see Fig.~\ref{fig:Fig1}. It should be noted that the minimum level of receptor's cross-reactivity (or maximum specificity) $(\alpha_\text{max})^{-1}$ defines a natural  scale against which we can measure  antigenic divergence $\delta$ and hence, form a dimensionless measure of antigenic divergence $\hat \delta \equiv \delta/(\alpha_\text{max})^{-1}$.

Immune-pathogen recognition depends  both  on the binding affinity $E_{\alpha,\theta}(r,\nu)$ and  the encounter rate $\gamma_\nu (t)$ between an immune receptor $r$ and the antigen  $\nu$ at a given time $t$. The encounter rate $\gamma_\nu (t)$ depends on the abundance of the antigen  and the immune receptor, and hence, can vary during an infection within a host.  The probability that a receptor $r$ encounters and binds to an antigen $\nu$   in a short  time interval $[t,t+dt]$ can be expressed by, $\rho (r,\nu,t) \text{d}t  =\gamma_\nu (t)  E_{\alpha,\theta}(r,\nu) \text{d}t$; a similar notion of encounter rate has been previously used in ref.~\cite{Mayer:2016jm}. A memory response in an individual is triggered through the recognition of an antigen  by a circulating memory receptor. If no such recognition occurs during the deliberation time $\tau\approx 1.5-5$ days, the immune system initiates a na\"ive response. Therefore, the probability that an antigen is recognized through a novel na\"ive response  $\P^{(0)}_{{}_\text{recog.}}$ can be expressed as the probability of the antigen not being recognized $1- \P^{(m)}_{{}_\text{recog.}}$ by an available memory receptor $r_m$ over the deliberation period $\tau$,
\EQA
\nonumber \P^{(0)}_\text{recog.}(\nu) &=& 1- \P^{(m)}_\text{recog.}(r_m,\nu)  \\
&=&e^{-\int_0^\tau \rho(\nu,t) \text{d} t  }  =e^{-E_{\alpha,\theta}(r_m,\nu)\Gamma(\nu,\tau)}\label{eq:recogProb}
 \EEA
where $\Gamma(\nu,\tau) = \int_0^\tau  \gamma_\nu (t) dt$ is the expected number   of pathogenic encounters  over the deliberation time $\tau$ and depends on the accumulated pathogenic load, as pathogens proliferate in the absence of an effective memory prior to a na\"ive response. Here, we have assumed that the affinity of  the memory receptor does not change over the response time, which is a simplification since memory receptor can undergo limited affinity maturation~\cite{Shlomchik:2020ab,McHeyzerWilliams:2018fv}. To further simplify, we also assume that the accumulated pathogenic load is independent of the type of the pathogen $\Gamma(\nu,\tau) \equiv \Gamma(\tau)$.  As pathogens evolve away from the primary infector, the binding affinity $E_{\alpha,\theta}(r_m,\nu)$ of the stored memory  receptor $r_m$, and hence, the  probability to mount a  memory response $\P^{(m)}_\text{recog.}(r_m,\nu,\tau)$ decays.  

The deliberation time prior to a novel response provides a window for memory to react with an antigen and mount an immune response by  initiating an irreversible cascade of downstream events.  Although initiation of this pathogenic recognition can be modeled as an equilibrium process, the resulting immune response is a non-equilibrium and an irreversible process, the details of which are not included in our model. \\

\section*{\large iii. Decision-making to mount a memory or na\"ive response}  

In the theory of decision-making, a rational decision-maker chooses between two possible actions $a \in \{\text{na\"ive, memory}\}$ each contributing  a utility $ U_a$. If the decision-maker has  prior preference for each action, which we denote by the prior probability distribution  $Q_0(a)$, its decisions could be swayed by this knowledge. As a result, the constrained decision-maker should choose actions according to an optimized probability density $Q(a)$, which maximizes the expected utility while satisfying constraints due to the prior assumption~\cite{vonNeumann:1944tc,Ortega:2013eu},
{\EQA
Q(a) = \underset{Q(a)}{\text{argmax}}\left ( \sum_{a}  U_a Q(a) - \frac{1}{\beta} D_{KL}\left(Q(a) || Q_0(a)\right) \right)
\EEA}
Here, {\small $D_{KL}(Q(a) || Q_0(a)) = \sum_a Q(a) \log \left(Q(a)/Q_0(a)\right) $} is the Kullback-Leibler distance between the  rational distribution $Q(a)$ and the prior distribution $Q_0(a)$ and  $1/\beta$ is a Lagrange multiplier that constrains the efficacy of a decision-maker to  process new information and deviate from its prior assumption. The optimal solution for a rational yet constrained decision follows,
\EQA 
 Q(a) = \frac{1}{Z} Q_0(a) e^{\beta  U_a}
 \label{eq:probUtility}\EEA
where $Z= \sum_a Q_0(a) e^{\beta  U_a}$ is a normalization factor. If information processing is highly efficient (i.e., the bias factor $1/\beta \to 0$) the rational decision-maker deterministically  chooses the action with the highest utility. On the other hand, if the prior is strong (i.e.,  $1/\beta \to \infty$), the decision-maker hardly changes its opinion and acts according to its prior belief (i.e., $Q(a)=Q_0(a)$). Moreover, if the prior distribution is uniform across actions (i.e., no prior preference), rational decision maximizes the entropy of the system~\cite{Jaynes:1957fy}, resulting in the probability of actions $Q(a) \sim \exp[\beta  U_a]$. In  our analysis, we consider the case of unbiased maximum entropy solution for decision-making. As a result the probability to utilize memory $Q_\text{mem.}$ or na\"ive $Q_\text{na\"ive}$ follows,
\EQA
Q_\text{mem.} = 1- Q_\text{na\"ive}= \frac{e^{\beta 	U_\text{mem} }}{e^{\beta 	U_\text{mem}} +e^{\beta 	U_\text{na\"ive} }}
\label{eq:memoryDecision_SI}\EEA
which is a sigmoidal function, dependent on the utility of each action. 

A decision to mount a memory or na\"ive response  $Q(a)$ based on their respective utilities (eq.~\ref{eq:probUtility})  should be consistent with the biophysical description of the immune response  through recognition of an antigen by either of these cell types (eq.~\ref{eq:recogProb}). By equating these two descriptions of an immune response  (eqs.~\ref{eq:recogProb},~\ref{eq:probUtility}) we can specify the utility gain associated with  mounting a memory  or a na\"ive response in terms of the biophysics and kinetics of receptor-antigen interactions,
\EQA
\nonumber Q_\text{mem.}  =\P^{(m)}_\text{recog.}(r_m,\nu)&\longrightarrow&  \frac{e^{\beta 	U_\text{mem} }}{e^{\beta 	U_\text{mem}} +e^{\beta 	U_\text{na\"ive} }}=1- e^{-E_{\alpha,\theta}(r_m,\nu)\Gamma(\nu,\tau)}\\
\nonumber &\longrightarrow& \beta (U_\text{mem.} - U_\text{na\"ive}) = \log \left[ e^{E_{\alpha,\theta}(r_m,\nu)\Gamma(\nu,\tau)} -1\right]\\
 \EEA
 Importantly, in the regime that memory  is efficient and being utilized to mount a response (i.e., a low chance for na\"ive recognition: $ P^{(0)}_\text{recog.} =e^{ - E(\nu) \Gamma(\nu,\tau)} \ll 1$),  the sigmoid form for  decision to use memory (eq.~\ref{eq:memoryDecision_SI}) is dominated by an exponential factor. Therefore, the utility gain by a memory or a na\"ive response to an evolved antigen  $\nu_i$ at an antigenic distance $d_i =\Vert\nu_i-\nu_0\Vert$ from the memory receptor's  cognate antigen $ \nu^*_{r_m} \equiv \nu_0$   follows (see Methods),
{\small
\EQA
\nonumber  U_\text{mem} (\Vert\nu_i-\nu_0\Vert;\alpha,\theta) &=& U_\text{na\"ive}+E_{\alpha,\theta}(r_m,\nu_i)  \\
\nonumber&=&-\Omega(\Gamma_\tau) +E_{\alpha,\theta}(\Vert\nu_i-\nu_0\Vert) \\
\EEA}
\noindent Here, we introduce the cost for deliberation $\Omega(\Gamma_\tau)$ as the negative utility of the n\"ive response  $U_\text{na\"ive}$. Deliberation cost $\Omega(\Gamma_\tau)$ is a  monotonically increasing function of the cumulative pathogen load $\Gamma_\tau$ and reflects the damage (cost) incurred by  pathogens as  they proliferate during the deliberation time $\tau$ prior to activation of the novel na\"ive response; see Fig.~\ref{fig:Fig1}. It is important to note that the  difference in the memory and the na\"ive utility {\small $\Delta U=  U_\text{mem}-  U_\text{na\"ive}$} determines the decision to mount either of these responses.

 The same consistency criteria between decision-making (eq.~\ref{eq:probUtility})  and cellular recognition (eq.~\ref{eq:recogProb})  indicates that the information processing factor  $\beta$ in eq.~\ref{eq:probUtility} should be equal to the accumulated pathogenic load  $\Gamma(\tau)$ during the deliberation period $\tau$: $\beta=\Gamma(\tau)$. A longer deliberation, which on one hand leads to the accumulation of pathogens, would allow the immune system to exploit the utility of a usable memory (i.e., process information), even if the memory has only a slight advantage over a responsive na\"ive receptor.  As a result, we refer to $\beta$ as the {\em deliberation factor.}  Moreover, this analogy relates the efficacy of information processing $\beta$, which plays the role of  inverse temperature in thermodynamics, and the total accumulated pathogenic load $\Gamma(\nu, \tau)$, which acts as the sample size for memory receptors as they encounter and accumulate information about pathogens. Interestingly, previous work has drawn a similar correspondence between the  inverse  temperature in thermodynamics and the effect of sample size on statistical inference~\cite{LaMont:2019es}. 

 The deliberation factor  in the immune system should be bounded $\beta \leq \beta_\text{max}$ in order for the organism to survive new infections by mounting a novel  response that can suppress an exponentially replicating pathogen before it overwhelms the host. Using this bound, we define rescaled deliberation factor $\hat \beta  \equiv\beta/\beta_\text{max}\leq 1$ to characterize the kinetics of an immune  response in a dimensionless fashion. 

{It should be noted that our decision-making formalism assumes that if memory is available, it can be utilized much more efficiently and robustly than a na\"ive response. Therefore, we do not consider scenarios where  memory and na\"ive responses are  equally involved in countering an infection--- a possibility  that could play a role in real immune responses. Nonetheless, since such mixed responses are relatively rare, we expect that including them in our model would only result in a slightly different interpretation of the  deliberation factor $\beta$ and should not qualitatively impact our results. 
 }

If the immune system decides to  mount a memory response against an evolved antigen $\nu_i$, the  binding profile of memory  against  the target pathogen remains unchanged and equal to the profile $E_{\alpha,\theta}(r_{\nu_0} ,\nu)$ against the primary infection ${\nu_0}$. However, if the immune system mounts a na\"ive response, a new memory receptor $r_{\nu_i}$ would be generated with a binding profile $E_{\alpha,\theta}(r_{\nu_i},\nu)$, centered around  the latest infection $\nu_i$. As a result, the expected binding profile $\overline{ E^{(i)}_{\alpha,\theta}}(\nu)$ at the $i^{th}$ round of infection  is an  interpolation between the profiles associated with memory and na\"ive response, weighted by the likelihood of each decision (eq.~\ref{eq:recogProb}), 
{ \small\EQA
\overline{ E^{(i)}_{\alpha,\theta}}(\nu)= \P^{(m)}_\text{recog.}(r_{\nu_0},\nu_i) E_{\alpha,\theta}(r_{\nu_0} ,\nu)+ \P^{(0)}_\text{recog.}(\nu_i)  E_{\alpha,\theta}(r_{\nu_i},\nu)
\label{eq:EffectiveProfile}
\EEA
}

 The expected binding profile at the $i^{th}$ round of infection $\overline{ E^{(i)}_{\alpha,\theta}}(\nu)$ (eq.~\ref{eq:EffectiveProfile}) deviates from the optimal profile centered around the infecting pathogen $E_{\alpha,\theta}(r_{\nu_i} ,\nu)$ (i.e., memory profile stored following a novel response); see Fig.~\ref{fig:Fig1}. This deviation arises because an energetically sub-optimal memory response can still be favorable when time is of an essence and the decision has to be made on the fly  with short deliberation. This tradeoff between the kinetics and the energetics of immune response results in a {\em non-equilibrium decision-making}~\cite{GrauMoya:2018gp} by the immune system. In analogy to non-equilibrium thermodynamics, we express this deviation as a dissipative cost of memory response $K_{\diss} (t_i;\alpha,\theta)$ at the $i^{th}$ round of infection (time point $t_i$), which we quantify by the Kullback-Leibler distance between the expected and the optimal binding profiles, in units of the deliberation factor $\beta$,

\EQA
\label{eq:Dissipaion_SI}
\nonumber K_{\diss} (t_i;\alpha,\theta) &=& \frac{1}{\beta} D_{KL}\left (\overline{ E^{(i)}_{\alpha,\theta}}(\nu) || E_{\alpha,\theta}(r_{\nu_i},\nu)\right) \\
\nonumber&=&\frac{1}{\beta}\sum_{\text{antigens: }\nu} \overline{ E^{(i)}_{\alpha,\theta}}(\nu) \log\left[\frac{\overline{ E^{(i)}_{\alpha,\theta}}(\nu) }{E_{\alpha,\theta}(r_{\nu_i},\nu)}\right]\\
\EEA
where we ensure that binding profiles are normalized over the space of antigens. The dissipation $K_{\diss}$ measures the sub-optimality (cost) of the mounted response through non-equilibrium decision-making and quantifies deviation from an equilibrium immune response~\cite{GrauMoya:2018gp}.

An optimal memory strategy should be chosen such that it maximizes the expected utility of the immune response $\left \langle  U\right \rangle =  U_\text{mem} \P^{(m)}_\text{recog.}+ U_\text{na\"ive}\P^{(0)}_\text{recog.}$, while minimizing  the dissipation cost due to the non-equilibrium response $K_{\diss}$, over the lifetime of an organism. To infer an optimal strategy, we introduce net utility $\U_\text{net}$ that  accounts for the  tradeoff between the expected utility and  dissipation at a given round of infection  at time point $t_i$,
\EQA
\label{eq:Unet_SI}
 \U_\text{net} (t_i;\alpha,\beta,\theta)=\left \langle  U_{\alpha,\beta,\theta}(t_i) \right\rangle  - K_{\diss}(t_i;\alpha,\theta)
\EEA

Net utility  can be interpreted as the extracted (information theoretical) work  of a  rational decision-maker that  acts in a limited time, and hence, is constantly kept out of equilibrium~\cite{GrauMoya:2018gp}.  We infer the  optimal memory protocol  (i.e., the optimal memory specificity $\alpha^*$ and deliberation factor $\beta^*$) by maximizing the total net utility of  memory responses throughout the lifetime of an organism, 
\EQA
 (\alpha^*,\beta^*)= \underset{\alpha,\beta}{\text{argmax} }\sum_{i:\text{ infections}} \U_\text{net} (t_i;\alpha,\beta,\theta).
\label{eq:Unet_opt_SI}
\EEA

{While we do not model time limits to memory, we effectively model only one memory at a time. This effect is the consequence of modeling the memory  as only being beneficial until a novel immune response is triggered resulting in the storage of an updated memory centered around a more recent antigen (Fig.~\ref{fig:Fig1}). After such an update, the old memory is no longer relevant as antigens have drifted away.

In our model, the characteristic time for a novel response (and memory update) is set by the expected antigenic divergence (Fig.~\ref{fig:Fig2}). Accordingly, cross-reactivity of memory is optimized so that the organism can mount effective responses against evolved forms of antigens in this window of time. However, if the  lifetime of memory were to be shorter than this characteristic time of memory update, we expect the organism to store more specific memory since this memory would be utilized to counter a more limited antigenic evolution before it is lost. In other words, the shorter of either the memory  lifetime or the characteristic time for memory updates determines the optimal cross-reactivity for immune memory.}
 }

\newpage{}

\appendix

\counterwithin{figure}{section}

  \setcounter{figure}{0}
 \renewcommand{\thefigure}{S\arabic{figure}}
  
\begin{figure}[h]
\centering
\includegraphics[width=\textwidth]{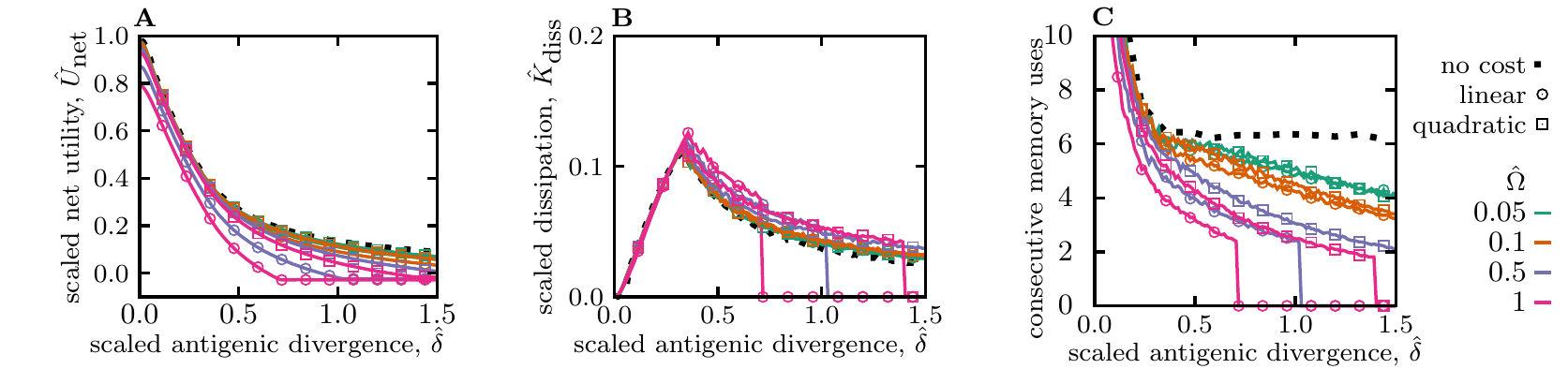}
\caption{\textbf{Utility, dissipation, and usage of optimal memory.}  {\bf (A)} and {\bf (B)}  show the scaled net utility $\hat U_\text{net} \equiv U_\text{net}/E_\text{max}$ (eq.~\ref{eq:Unet_SI}) and dissipation $\hat K_\text{diss} \equiv K_\text{diss}/E_\text{max}$ (eq.~\ref{eq:Dissipaion_SI}) per round of infection as a function of the antigenic divergence  $\hat\delta$. Rescaling by $E_\text{max}$ sets the magnitude of net utility for a response to conserved antigens (with $\hat \delta =0$), and in the limit of zero  deliberation cost $\hat{\Omega}\to 0$, to 1;  see Fig.~\ref{fig:Fig2} in the main text for comparison. {\bf (C)} The expected number of rounds that a  memory receptor is utilized prior to a novel response in an optimal system is shown to decay as the antigenic divergence $\hat \delta$ increases. The results are evaluated for  immune systems with optimized strategies ($\hat \alpha^*$,  $\hat \beta^*$) against pathogens with a given scaled antigenic divergence $\hat \delta$; the corresponding strategies are  shown in Fig.~\ref{fig:Fig2}.  Colors / markers indicate different na\"ive cost functions for deliberation, including no-cost $\hat \Omega\equiv\Omega/E_\text{max} =0$, linear cost $\hat \Omega  =\hat \Omega_0 \hat \beta$, and quadratic cost $\hat \Omega= \hat \Omega_0 \hat \beta^2$, with varying amplitudes $\hat\Omega_0$.  Simulation parameters: $\alpha_{\text{max}}=4$, $\beta_{\text{max}}=10$, and $\theta=2$.}
\label{fig:2sf1}
\end{figure}

\vspace{3cm}
 \begin{figure}[h]
\centering
\includegraphics[width=\textwidth]{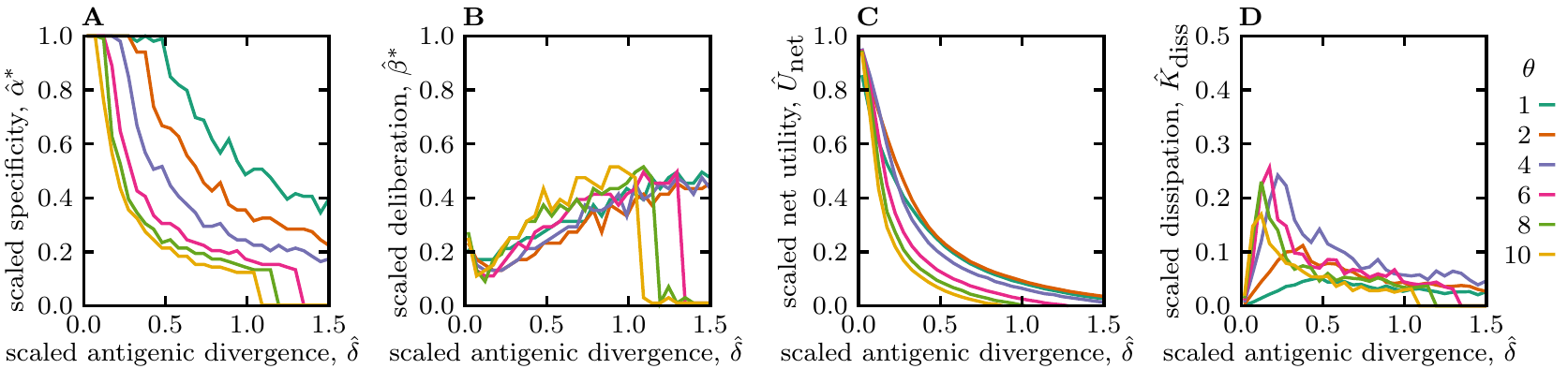}
\caption{{\bf Optimal memory strategies for different specificity shape factors $\theta$.} {\bf (A)} Scaled specificity $\hat \alpha^* \equiv \alpha^*/\alpha_\text{max}$, {\bf (B)} scaled deliberation factor $\hat \beta^*\equiv\beta^* /\beta_\text{max}$, {\bf(C)} scaled net utility $\hat U_\text{net} \equiv U_\text{net}/E_\text{max}$, and {\bf (D)} scaled dissipation are shown as a function of the scaled antigenic divergence per infection  $\hat\delta = \delta /(\alpha_\text{max}^{-1})$ (similar to Fig.~\ref{fig:Fig2}). Colors indicate different shape factors $\theta$ of the specificity  function, ranging from a double-sided exponential ($\theta= 1$), to Gaussian for $\theta =2 $ (as in Fig.~\ref{fig:Fig2}), and  top-hat functions $\theta >2$. The  dependence of memory strategies on antigenic divergence is qualitatively insensitive to the shape factor of the specificity function. Simulation parameters: linear deliberation cost function $\Omega= \hat\Omega_0 \hat \beta$ with $\hat \Omega_0 = 0.1$, $\alpha_{\text{max}}=4$,  and $\beta_{\text{max}}=10$.}
\label{fig:2sf2}
\end{figure}

\begin{figure}[h]
\centering
\includegraphics[width=\textwidth]{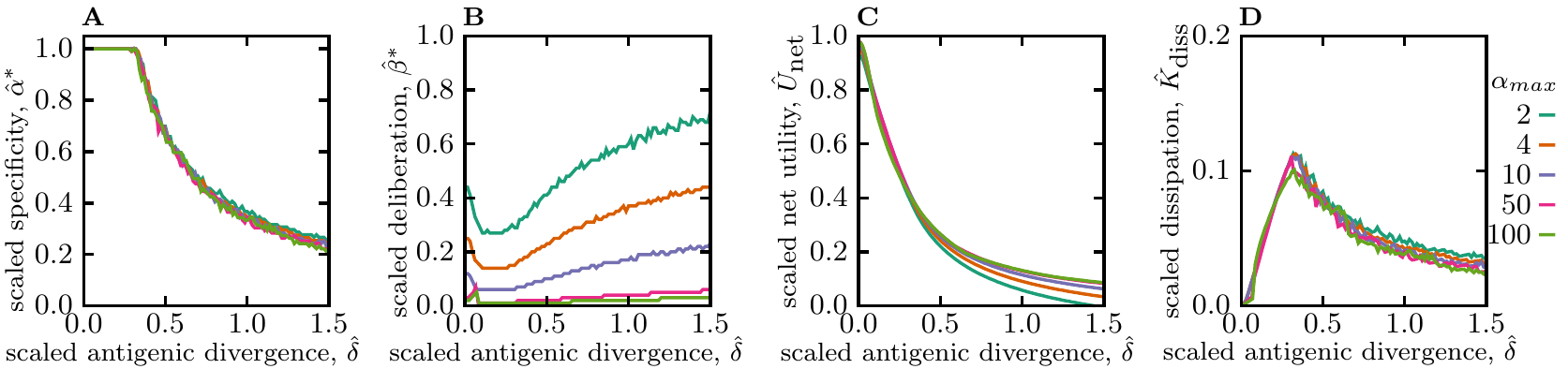}
\caption{{\bf Optimal memory strategies for different specificity thresholds $\alpha_\text{max}$.}  {\bf (A)} Scaled specificity $\hat \alpha^* \equiv \alpha^*/\alpha_\text{max}$, {\bf (B)} scaled deliberation factor $\hat \beta^*\equiv\beta^* /\beta_\text{max}$, {\bf(C)} scaled net utility $\hat U_\text{net} \equiv U_\text{net}/E_\text{max}$, and {\bf (D)} scaled dissipation are shown as a function of the scaled antigenic divergence per infection  $\hat\delta = \delta /(\alpha_\text{max}^{-1})$ (similar to Fig.~\ref{fig:Fig2}).  Colors indicate different  specificity thresholds $\alpha_\text{max}$. Memory strategies are qualitatively insensitive to the specificity threshold. Simulation parameters: linear deliberation cost function $\Omega= \hat \Omega_0 \hat \beta$ with $\hat \Omega_0 = 0.1$ and $\beta_\text{max}=10$.}
\label{fig:2sf3}
\end{figure}

\begin{figure}[h]
\centering
\includegraphics[]{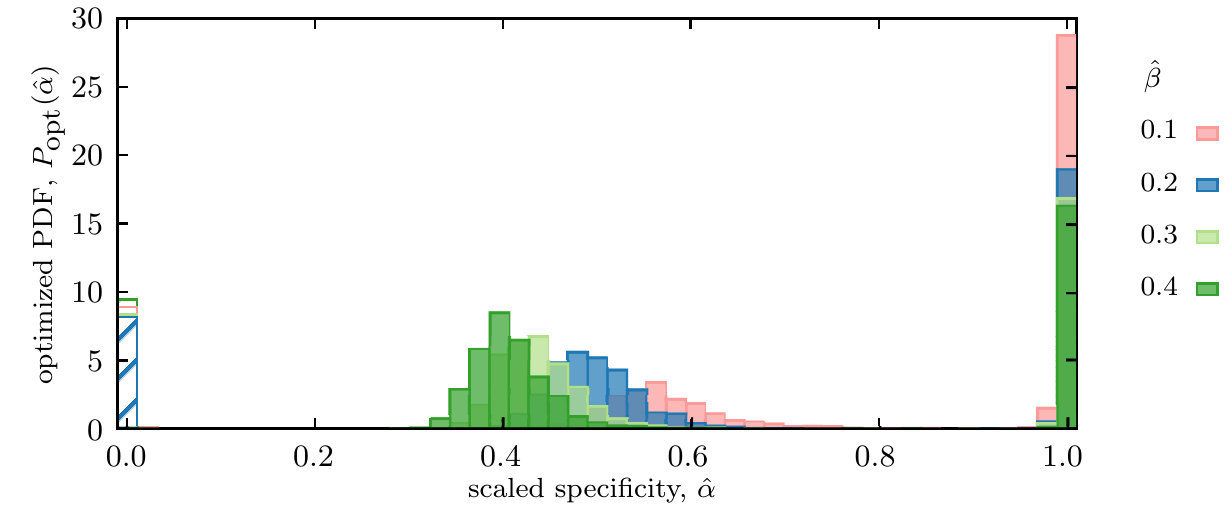}
\caption{{\bf Mixed memory strategy against pathogens for different deliberation factors $\hat \beta$.}  Distribution of  scaled optimized specificities $\hat\alpha^*$ of functional memories is shown for an immune system with a fixed deliberation factor $\hat \beta = 0.2$, in which  a mixture strategy with a bimodal distribution of specificities  $P(\hat \alpha)$ is established to counter  pathogens with a broad range of antigenic divergences, drawn uniformly  from a range $ \hat \delta \in (0,1.6)$ (similar to Fig.~\ref{fig:Fig3}). The dashed bars indicate stored memory with specificity $\alpha=0$, which is not further used in response to infections. Colors indicate different deliberation factors. Simulation parameters:  $\alpha_\text{max}=4$, and $\beta_\text{max}=10$. }
\label{fig:3sf1}
\end{figure}

\begin{figure}[h]
\centering
\includegraphics[]{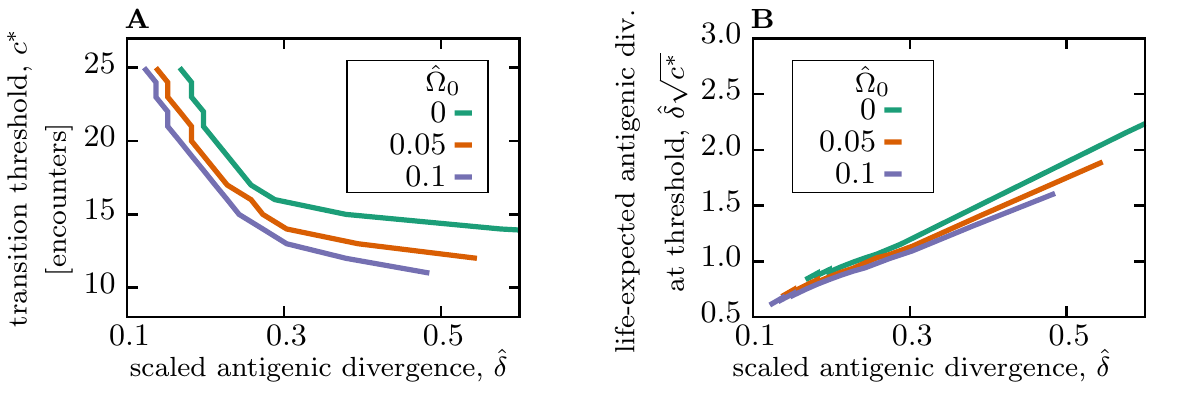}
\caption{ {\bf Pathogen encounter threshold to transition between cross-reactive and specific  memory.} {\bf (A)} The encounter threshold  $c^*$, shown in Fig.~\ref{fig:Fig4}A,B, decays as a function of  the antigenic divergence (per encounter) $\hat \delta$ and the amplitude of the naive cost $\hat \Omega_0$ (colors). {\bf(B)} The expected antigenic divergence for the duration of $c^*$ (threshold) encounters  $\hat{\delta} \sqrt{c^*}$ is shown as a function of antigenic divergence (per encounter) $\hat \delta$. Simulation parameters: linear deliberation cost function $\Omega= \Omega_0 \hat \beta$, $\alpha_{\text{max}}=4$, $\beta_{\text{max}}=10$, and $\theta=2$.}
\label{fig:4sf1}
\end{figure}

 \end{document}